\documentclass[acmsmall]{acmart}

\usepackage{booktabs} 
\usepackage{graphicx}
\usepackage{makecell}
\usepackage{subcaption}
\usepackage{caption}
\usepackage{tabularx}
\usepackage{color}
\usepackage{hyperref}
\usepackage{array}

\usepackage[ruled]{algorithm2e} 

\SetAlFnt{\small}
\SetAlCapFnt{\small}
\SetAlCapNameFnt{\small}
\SetAlCapHSkip{0pt}
\IncMargin{-\parindent}

\copyrightyear{2020}

\title[Gender Diversity in CS]{Gender Diversity in Computer Science at a Large Research University}  
\thanks{This work is an extension of the work published as \textit{Exploring Gender Diversity in CS at a Large Public R1 Research University} at a SIGCSE Conference.}

\author{Monica Babe\c{s}-Vroman}
\affiliation{%
  \institution{Computer Science Department, Rutgers University}
}
\email{monica.babes@cs.rutgers.edu}

\author{Thu D. Nguyen}
\affiliation{%
  \institution{Computer Science Department, Rutgers University}
}
\email{tdnguyen@cs.rutgers.edu}

\begin{abstract}
With the number of Computer Science (CS) jobs on the rise, there is a
  greater need for Computer Science graduates than ever.
  At the same time, most CS departments across the country are only seeing $25$-$30$\% of
  female students in their classes, meaning that we are failing to
  draw interest from a large portion of the population.
  In this work, we explore the gender gap in CS at a large public research university, using three data sets that
  span thousands of students across 5.5 academic years.  By combining
  these data sets, we can explore many issues such as
  retention as students progress through the CS major. For example,
  we find that a large percentage of women taking the Introductory
  CS1 course for majors do not intend to major in CS, which contributes to a
  large increase in the gender gap immediately after CS1.  This
  finding implies that a large part of the retention task is
  attracting these women to further explore the major.
  We report findings in three areas of research in the context of the CS department at our university: the CS environment, the computing background of our students, and the students' grades.
  These findings may also be applicable to computing programs at other large public research
  universities.

\end{abstract}

\begin{document}

\maketitle

\renewcommand{\shortauthors}{M. Vroman and T.D. Nguyen}

\section{Introduction}
\label{s:intro}

The need for computing majors in the workforce is greater than
ever. The U.S. department of labor estimates that by 2026 there will be nearly $3.5$ million computing related jobs available with enough graduates for only $17\%$ of them at current graduation rates \cite{ncwit19}.  The deficit in Computer Science (CS) graduates is partly due to the fact that CS is not widely taught in middle school and high school and that colleges do not generate the number of graduates needed by the workforce.

Over the last several years, there has been an increase in enrollments
in CS departments across the country, although they are still not
enough to cover the estimated number of jobs.  From a diversity
perspective, only about $25$-$30$\% of enrolled students are women (and
even fewer are ethnic minorities) \cite{nsf12}. Increasing the number
of women with Computer Science degrees is critical in providing the needed
workforce in computing and, at the same time, increasing diversity which is crucial in the creation of technology~\cite{margolis03}. Environments that are gender diverse perform better, have more innovation and productivity, and have a more supportive infrastructure \cite{ncwit14}. 

The gender gap in college has been extensively studied~\cite{sax08}.  Linda Sax has pointed out that it is important to study each
discipline separately~\cite{sax12}, rather than looking at all STEM
disciplines together as in previous work~\cite{sax08}, since root causes behind gender gaps can be different among different disciplines.  Along these lines, there have also been multiple studies of the gender gap in CS~\cite{alvarado10,margolis03,vilner06}. 
In this work, we add to this body of knowledge using extensive
data sets from a large public research university.  We believe that
this data and the accompanying analyses are valuable
because large departments at institutions similar to ours generate a
considerable percentage of the computing workforce in the country.

Specifically, in this paper, we analyze student data from a set of
four core courses that all majors in our undergraduate CS program are
required to take.  Our data comprises three different data sets:
one data set contains demographic data, course information, and
grades; the second one comes from an Introductory Survey 
our CS1 students take and contains information about each student's
computing background and how likely they are to pursue the CS major
(among other information); the third data set comes from an Exit
Survey, asking CS1 students, among other information, about the usefulness of resources that were available to them in CS1 and how likely they were to major in
Computer Science before and after taking CS1. We describe these data sets in more detail in Section \ref{s:data}.

Using these data sets, we answer research questions in the following categories:

\begin{itemize}
\item \textbf{The CS Environment:} In Section~\ref{s:environment}, we explore topics such as whether or not there is a gender gap in CS at our university, how the gender gap has changed over the last few years, and how it changes from introductory classes to advanced classes in the major. We also answer the question, at what point in the major is the loss in the proportion of women of women the greatest? We also explore gender differences in intent to major, comfort level in using various resources, interaction with peers, etc.
\item \textbf{Students' Background:}  In Section~\ref{s:background}, we answer questions about our students' prior computing experience and how prior experience correlates with retention, as well as gender differences in the use of technology, whether or not students received encouragement to pursue CS, and how students assess their own computing related interests and abilities.  
\item \textbf{Students' Grades:} In Section~\ref{s:grades}, we explore correlations between the grades received in CS1 and a few factors such as the decision to take CS2 for men and women, a change in intent to major after taking CS1, familiarity with Java, and the students' self-assessment of various CS related interests and abilities.
\end{itemize}

A main contribution of our work is the combining of three data
sets to answer a number of questions about our CS student body that, as far as we
know, have not been answered before.  For example, we give concrete
data on where along a path of four required courses women decide to leave the CS program, and show the retention rates for men and women 
who take the CS1 course intending to major in CS compared to those not
intending to major in CS.  Additional results match current knowledge
and accepted wisdom, but we provide concrete numbers from a large public research institution.

Many universities across the country are making efforts to diversify their student population in computing. Various initiatives\cite{klawe13, margolis03} have successfully led to classes that are more diverse, some with up to $40$-$50$\% women. 
In our future work, we are planning to implement some of these initiatives and measure their impacts.

\section{Background and Related Work}
\label{s:related}

A number of studies \cite{alvarado10, vilner06} have looked at
computing students' enrollment data and analyzed gender differences in
enrollment numbers and pass/fail rates in CS classes. Our work
  integrates data on enrollments, grades, and surveys to dig deeper and
answer questions about intent to major, how many students actually
change their mind about majoring, and what are our actual retention
rates. Other related papers report on survey data \cite{cohoon08,
  frieze13} or on data from interviews \cite{falkner15, habib14,
  margolis03} and assess the female students' attitudes, motivations
and confidence in computing. We use both surveys and enrollment data
to link our observations on gender differences with grades.

One paper analyzes students' grades on seven projects in an introductory programming course \cite{sahami16} and fits these grades to a mixture model with two gaussian distributions. While this work also analyzes student data, it is mainly focused on students' grades and the analysis does not include any other student information.

A few other papers have analyzed student data with the goal of understanding phenomena such as gender gaps in college courses. The freshmen survey \cite{heri} provided one of the largest such databases. Their data has been extensively analyzed \cite{sax08} and has answered questions on how men and women attending college are different in terms of background, achievement, perceptions of their environment, etc. Our work focuses on gender differences in computer science specifically. 

Previous work has also addressed the issue of low female representation in STEM disciplines \cite{shapiro11}, looking at educational factors that influence this phenomenon and making suggestions on how to change this trend. In computer science specifically, many have asked the question why are there so few women majors \cite{beyer14, cohoon01, cohoon08, falkner15, habib14, margolis03} and strategies to close the gender gap have been proposed \cite{alvarado10, cohoon02,frieze13, goode08, knobelsdorf07, pivkina09, sax15,
scutt13, rheingans18, ibe18}, including addressing the attitudes of students toward computer science ~\cite{alvarado17, wang17, blaney17, lishinski16, rorrer18}.
Our work does not directly address the reasons behind low female representation in CS, but  rather uses student data from a large research university to explore the relationship between gender and a variety of factors including students' computing background, how much they are helped by the resources offered, etc.

A number of computing departments at North American universities have made it their goal to increase the percentage of women in their classrooms. Some of their initiatives included changing their CS1 class to include more real-life applications \cite{guzdial14, klawe13}, offering learning opportunities to students who did not have prior experience \cite{berkeley, klawe13}, providing research projects for undergraduate female students \cite{berkeley, klawe13}, building a solid community of women in computing \cite{klawe13, margolis03}, engaging faculty in recruitment \cite{berkeley} and training them on how to design engaging classes \cite{guzdial14}, increasing the diversity of the faculty \cite{berkeley}, and reaching out to middle schools and high-schools \cite{berkeley, guzdial14}.

Our data and analysis may aid these and other universities in their efforts to narrow the gender gap in CS.

\section{Methodology}
\label{s:data}

This research was conducted at a large public research institution in the United States.  Our student body is made up of more than $50,000$ undergraduate
students from all $50$ states and more than $100$ countries. The
percentage of women is about $50$\%, and the percentage of students who self-identify as non-white is about $58$\%.
 
Our study uses three data sets. All the data is anonymized, but allows entries to be linked between data sets by anonymized student ids. 

The first data set focuses on student demographics (gender, graduation
year, and ethnicity) and grades. We call this data
set the Registrar data. It allows us to track students taking a
sequence of four classes as they make their way through the CS
major. These classes broadly cover foundational CS concepts, including
Introduction to Programming (CS1), Data Structures (CS2), Computer Architecture (CS3), and Algorithms (CS4).  All four are required for the undergraduate CS major.

The first three classes, CS1, CS2, and CS3, form a direct sequence,
with CS3 requiring CS2 as a prerequisite, and CS2 requiring CS1.  The
4th class, CS4, requires CS2 as a prerequisite, and so may not always
be taken after CS3.  However, it is the highest level class required
for the major, and the vast majority of our students delay CS4 until
after CS3 (often by several semesters).  Thus, the students' progression
through this sequence of courses is very indicative of their
progression through the major.

The Registrar data contains all the students who have taken any
classes in our department during the fall and spring semesters between Fall 2012 and Fall 2017 ($n=51,494$) and, therefore, all the students in the other two data
sets. 

The second data set, the Introductory Survey, comes from surveys taken
at the beginning of CS1 during every Spring and Fall semester between Fall 2012 and Fall 2017, except for Fall 2014 and Spring 2015. Each survey asks students about their demographic information (age and gender), what is the
students' tentative or declared major, what kind of prior programming
experience they have (with the options: high-school advanced
placement, self-taught, other college course, and none), and what is
the probability they will continue in CS (with options $0$\%, $25$\%,
$50$\%, $75$\%, and $100$\% or ``Very unlikely'', ``Unlikely'', ``Neutral'', ``Likely'', and ``Very likely'', depending on the semester).  This data set contains answers from a
subset of students taking the introductory class CS1 ($n=3,728$).

The third data set, the Exit Survey, comes from another optional survey taken at the
end of our CS1 class during every fall and spring semester between Fall 2015 and Fall 2017 ($n=2,748$). It asks
students how likely they were to major in CS before and after taking
the class, how many lectures they attended, how many hours per week
they spent on the class on average, how often they used the resources
offered, and how helpful they found these resources to be. Many of
these survey questions were designed to collect information that is
outside the scope of this study.

Our data sets do not include information about students who have taken any of the CS courses during the summer semesters. This fact may introduce some errors in studying the student population in these courses.

The above data sets complement each other to give a more comprehensive
picture of who our students are, what are their backgrounds, what
classes they take, and how the CS1 class impacts them.  When analyzing
our data, we use the chi-square test for statistical significance to
detect significant differences between groups. In the remainder of the
paper, we always explicitly point out whenever we talk about
differences that are not statistically significant.  All other
differences discussed are statistically significant.

\section{The CS Environment}
\label{s:environment}

\subsection{Gender Gap in CS}
\label{ss:gender_gap}

\begin{figure}[!htb]
\minipage{0.38\textwidth}
 \includegraphics[width=\textwidth]{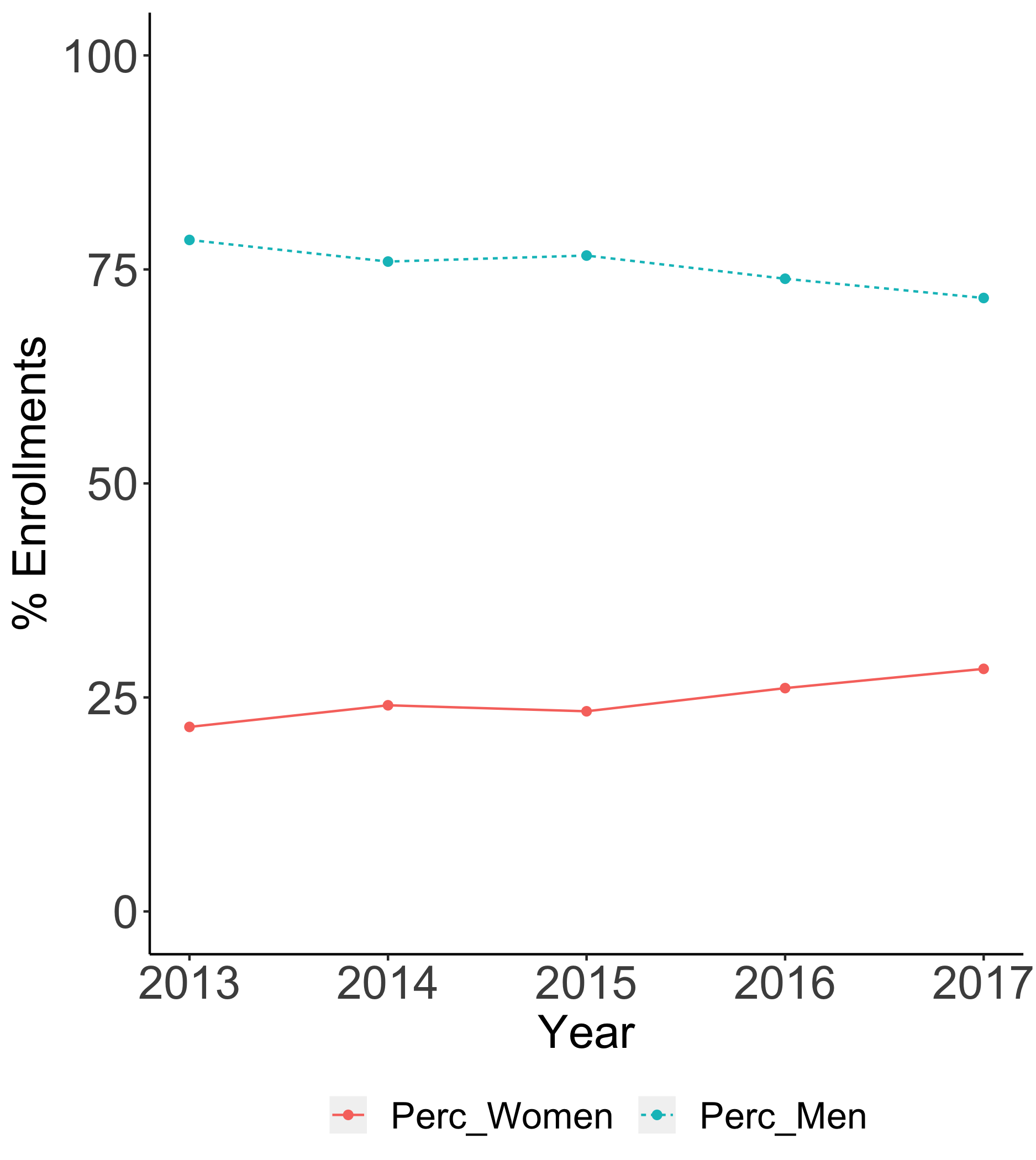}
  \caption{Percentage of Men and Women in CS1.}
  \label{f:enrollments_sem_comb}
\endminipage\hfill
\minipage{0.58\textwidth}%
  \includegraphics[width=\textwidth]{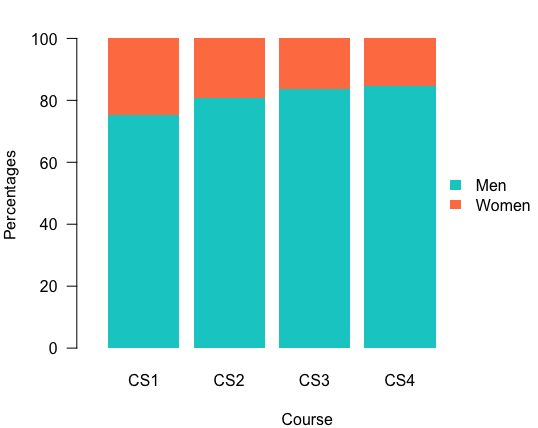}
  \caption{Percentage of Men and Women in CS1 - CS4.} 
  \label{f:perc_men_women_cs1_cs4}
\endminipage
\end{figure}

We begin this section with answering a basic question: is there a gender gap
is our CS classes?  Figure~\ref{f:enrollments_sem_comb} shows the percentages of men and women in CS1
over the past five years.  Figure~\ref{f:perc_men_women_cs1_cs4} shows the percentages of men and women in CS1 through CS4  accumulated over the past five years.  Both are derived from the Registrar
data.  Clearly, there is a significant gender gap that is close to
national averages \cite{nsf17}.

The data also answers three additional important questions.  First, is
our gender gap growing over time, especially with the recent rapid
rise in enrollments?  Encouragingly, Figure~ \ref{f:enrollments_sem_comb} shows
that \textit{the gap has been narrowing, although slowly, in recent years.} 

For the numbers used to generate Figure~\ref{f:enrollments_sem_comb}, the difference in the proportion of men and women between 2013 and 2016, 2015 and 2017, and 2013 and 2017 are statistically significant  according to the chi-square test ($p<0.05$). These observations
are in contrast to findings reported during the last boom in CS
enrollment, when the number of women entering college interested
in the CS major grew more slowly than the number of men~\cite{sax-ncge-2014}.

Second, is the gender gap growing as students progress toward
graduation?  We can see (Figure~\ref{f:perc_men_women_cs1_cs4}) that the gender gap does indeed grow as
students move toward graduation, starting with almost $25$\% female
students in CS1 and dropping to about $15$\% in CS4. Interestingly, even though the percentage of women enrolled in CS1 has been slightly increasing in recent years (Figure~\ref{f:enrollments_sem_comb}), our data shows that the percentage of women in CS4 has not had a statistically significant increase during the same period of time.

Finally, when do women drop out of the CS undergraduate program?  The
statistically significant drops in the participation of women are between
CS1 and CS2 and between CS2 and CS3. Previous research ~\cite{babes17} showed a significant drop in the percentage of women only between CS1 and CS2, whereas the drop between CS2 and CS3 was not statistically significant at that time. In our data, the drop between CS3 and CS4 is smaller, and also not statistically significant.
We conclude that the most dramatic drops in the percentage of women are after CS1 and CS2, although there may be a trend of further smaller losses as students progress toward graduation.  This observation matches conventional wisdom, but our data provides concrete numbers from a large public university.

\subsection{Gender Gaps in Courses}

Are there different gender gaps in different CS courses? To answer this question, we use the Registrar data set. We combine the data on students from all the semesters between Fall 2012 and Fall 2017 and count how many men and women have been enrolled in each course during that time period. 

Courses with the biggest gender gaps have a percentage of women varying from \textbf{12\%} to \textbf{16\%} and the difference in the proportion of men and women among these courses is not statistically significant. These are courses such as Introduction to Discrete Structures, Software Methodology, Systems Programming, Principles of Programming Languages, Images and Multimedia, Design and Analysis of Algorithms, Internet Technology, Compilers, Design of Operating Systems, Distributed Systems, Computer Security, Introduction to Computer Graphics, Software Engineering, and Introduction to Artificial Intelligence\footnote{\small Some of the names of these course have been changed to generic names for anonymity purposes.}. The smallest gender gap is in a non-majors course, computer applications for business. Among courses for majors, the smallest gender gap is between \textbf{18\%} women and \textbf{25\%} women , with no statistically significant difference in the proportion of men and women among these courses. Courses in this category are Introduction to Computer Science, Formal Languages and Automata, and Introduction to Computer Robotics.

The difference in the proportion of men and women between any course in the first set of courses and any course in the second set of courses is statistically significant.

\subsection{Level of Interest in Majoring and Correlation between Intent to Major and Continuation Rates.}

The observation that the highest drop in the percentage of women is after CS1 and then after CS2 (Section \ref{ss:gender_gap}) raises an important but unexpected question: is there a
difference in the level of interest for majoring in CS between the men
and women taking CS1?  We did not expect this question to be a factor
when we began this study because our CS1 class is intended for CS
majors and has a reputation for being difficult; we
have several non-major courses with very large annual enrollments.
However, we did ask our students about their intended major in our
Introductory Survey for completeness.

The survey had slight variations between semesters, but, generally included two questions, with the first asking what was the students' intended or declared major and the second asking students to estimate how likely they were to choose to major in  CS. A student was considered as
intending to major in CS if he/she had either answered ``Computer
Science'' to the first question or ``likely'' or ``highly likely'' to the second question or both.  This criterion offered the broadest interpretation for intention to major in CS.

Using the above formula and Introductory Survey data, we found that only about $59$\% of the female students in CS1 intended to major in CS before taking the class compared to about $76$\% of the male students, and the difference in these proportions was statistically significant.  Both of these numbers were surprising to us!\footnote{\small In retrospect, perhaps we should not have been surprised.  CS1 can be used to meet parts of the requirements for all students regardless of major.
  Many students seem to be taking the more rigorous CS1 course rather
  than the non-major courses toward this purpose. CS1 can also be substituted for introductory programming courses in other majors.}
Tables~\ref{t:indending_major_cs2} through
\ref{t:not_indending_major}, derived by correlating the Introductory
Survey data and the Registrar data, show the impact of these numbers.

The numbers in Tables~\ref{t:indending_major_cs2} and
\ref{t:not_indending_major_cs2} include students who completed the Introductory Survey during Fall 2012
through Spring 2014, and Fall 2015 through Fall 2016, while those in
Tables~\ref{t:indending_major} and \ref{t:not_indending_major} include
only students who completed the Introductory Survey during Fall 2012 through Fall 2015,
(since students need a longer length of time to
get to CS4).

\begin{table}[h]
\centering
\caption{Proportions of CS1 students who intend to major in CS. The difference in the proportion of men and women who go on to take CS2 is statistically significant.}
\begin{tabular}{ccc} \hline
&Take CS2& Do not take CS2\\ \hline
Female & $252$ ($63.8$\%)& $143$ ($36.2$\%) \\ \hline
Male & $512$ ($54.2$\%) & $433$ ($45.8$\%) \\ \hline
\end{tabular}
\label{t:indending_major_cs2}
\end{table}

\begin{table}[h]
\centering
\caption{Proportions of CS1 students who do not intend to major in CS. The difference in the proportion of men and women who take CS2 is not significant here.}
\begin{tabular}{crr} \hline
&Take CS2&Do not take CS2\\ \hline
Female & $81$ ($27.4$\%)& $215$ ($72.6$\%) \\ \hline
Male & $149$ ($31.8$\%) & $320$ ($68.2$\%) \\ \hline
\end{tabular}
\label{t:not_indending_major_cs2}
\vspace{-0.15in}
\end{table}

\begin{table}[h]
\centering
\caption{Proportions of CS1 students who intend to major in CS. The difference between men and women in the proportion of students who take CS4 is statistically significant.}
\begin{tabular}{crr} \hline
&Take CS4& Do not take CS4\\ \hline
Female & $61$ ($28.8$\%)& $151$ ($71.2$\%) \\ \hline
Male & $290$ ($38.9$\%) & $456$ ($61.1$\%) \\ \hline
\end{tabular}
\label{t:indending_major}
\end{table}

\begin{table}[h]
\centering
\caption{Proportions of CS1 students who do not intend to major in CS. The difference between men and women in the proportion of students who take CS4 is not statistically significant here.}
\begin{tabular}{crr} \hline
&Take CS4&Do not take CS4\\ \hline
Female & $24$ ($12.9$\%)& $162$ ($87.1$\%) \\ \hline
 Male & $31$ ($11.8$\%) & $231$ ($88.2$\%) \\ \hline
\end{tabular}
\label{t:not_indending_major}
\vspace{-0.15in}
\end{table}

Clearly, in terms of progressing from CS1 to CS2, the percentages are
higher for students who intend to major in CS than for students who do
not intend to major in CS. When this fact is coupled with the high
percentage of women taking CS1 but not intending to major in CS, we
find a possible major factor for the loss of women from CS1 to CS2.
{\em The gender gap between men and women intending to major in CS is larger than
  the gap between men and women taking CS1.}  That is, many women
choose to take CS1 despite the fact that they do not intend to major in
CS, and we do not successfully attract them to the major.  This
matches the known fact that women often lose interest in CS before
they reach college level.  Yet, it also points to an opportunity: if
we can (re)kindle interest for the CS major in these women who choose
to take CS1, we have a chance to reduce the current loss of women after
CS1.

Interestingly, Table~\ref{t:not_indending_major_cs2} shows that a
non-trivial percentage ($27.4\%$) of the women who do not intend to major
in CS go on to take CS2.  This implies that while we still need to
consider how to improve CS1 to encourage more women (in fact, to
encourage more students of both genders) to further explore the CS
major, at least not all women not intending to major in CS are lost
immediately following CS1.  We currently do not know whether the
students are taking CS2 because they became interested in the CS major
or for other reasons.  We plan to survey CS2 students in the near
future.  Unfortunately, Table~\ref{t:not_indending_major} shows that
by CS4, most of the women who did not intend to major in CS at the beginning of CS1 have actually chosen to not continue with the
major.  This implies that there is work to be done in classes beyond CS1.

Finally, Table~\ref{t:indending_major} shows that the percentage of men intending to major in CS taking
CS1 and persisting to CS4 is higher than that of women. The differences in Table~\ref{t:not_indending_major} are not statistically significant.  Comparing the data in Tables~\ref{t:indending_major_cs2} and \ref{t:indending_major}, we see an inversion in the order of the percentage of men and women taking CS2 and CS4, respectively: for CS1 students intending to major in CS, the percentage of women taking CS2 is higher than the percentage of men taking CS, while the percentage of women taking CS4 is lower than the percentage of men taking CS4. This fact suggests that, by CS4, many women have lost their interest in the CS major. Existing literature points to possible causes for this phenomenon~\cite{beyer14, cohoon01} and we are planning on exploring it in more depth in our future work.

\subsection{Change in Intent to Major}

How does intent to major change from the beginning of the semester to the end of the semester in the CS1 class for men and women? To answer this question, we use the Exit Survey from the following semesters: Spring 2016, Fall 2016, Spring 2017, and Fall 2017. These surveys were worded differently during different semesters, but, generally, they asked two questions related to intent to major before and after the class. From answers to these questions, we computed the number of men and women who wanted to major in CS both before and after taking CS1 (yes to yes), wanted to major before taking the class and decided to not major after taking the class (yes to no), did not want to major before taking the class and decided they wanted to major after taking the class (no to yes), and did not want to major either before or after taking CS1 (no to no). We show these numbers in Table~\ref{t:intent_major}. The difference in the percentage of students in each of these categories is significant between men and women ($p< 2.2e-16$).

\begin{table}[h]
\centering
\caption{Intent to Major before and after Taking CS1 for Women and Men. ``yes to yes'' means that the students in this category intended to major before and after taking the class.}
\begin{tabular}{crrrr} \hline
&Yes to Yes&Yes to No&No to Yes&No to No\\ \hline
Female & $305$ ($69.4$\%)& $54$ ($12.3$\%)&$21$ ($4.8$\%)& $59$ ($13.4$\%) \\ \hline
Male & $1057$ ($80.7$\%) & $123$ ($9.4$\%)& $40$ ($3.0$\%) & $90$ ($6.9$\%) \\ \hline
\end{tabular}
\label{t:intent_major}
\vspace{-0.15in}
\end{table}  

Similarly to the numbers in the previous section, the numbers in Table~\ref{t:intent_major} point to men's stronger enthusiasm for majoring in CS, compared to the women which has been reported before\cite{cohoon08, margolis03}.

\subsection{Level of Comfort in Approaching Instructors, Peer-Leaders, etc}

In our CS1 class, students have access to multiple resources if they need help understanding the course material or if they need help with their homework and  programming assignments. Instructors teach the lectures and are available to answer questions during office hours. Peer leaders are undergraduate students who have taken the same classes in the past and done well. They conduct small recitations focused on problem solving. Tutors are available to answer students' questions in a small community space.

\begin{figure}
  \begin{minipage}[c]{0.67\textwidth}
    \includegraphics[width=\textwidth]{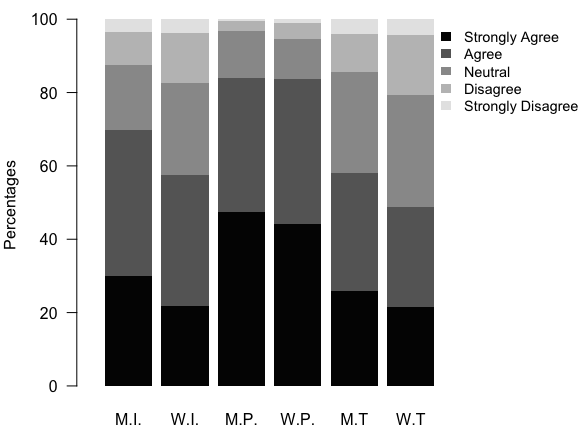}
  \end{minipage}\hfill
  \begin{minipage}[c]{0.3\textwidth}
    \caption{Level of agreement with ``I feel comfortable approaching''  instructors for men (M.I.) and women (W.I.), peer-leaders for men (M.P.) and women (W.P.), and tutors for men (M.T) and women (W.T.) with questions.} \label{f:comfort_levels}
  \end{minipage}
\end{figure}

To assess whether there are gender differences in how comfortable students feel in approaching instructors, peer-leaders, and tutors with questions, we use Exit Survey answers from Spring 2017 and Fall 2017, and Registrar data. The Exit Survey given during Spring 2017 and again during Fall 2017 asked three questions with answers in the 5 point Likert scale format, asking students how much they agreed or disagreed with the following statements: ``I feel comfortable approaching the instructor with questions'', ``I feel comfortable approaching the peer-leader with questions'', and ``I feel comfortable asking the tutors at the [community space] questions''.  We used the Registrar data to get gender information about the students and show the results in Figure~\ref{f:comfort_levels}. We tested the differences between the answers for men and women for all three groups (instructor, peer leader, and tutor), and found that the only statistically significant difference ($p<0.05$) was for the comfort level in asking the instructor questions. Our numbers show that women feel less comfortable than men when approaching the instructor, a result that has been discussed in the past\cite{margolis03, shapiro11}.

Is there a difference in level of comfort approaching the instructor, peer leader, and tutor between students who are more likely to major in CS after taking CS1 and students who are less likely to major in CS after taking CS1? To answer this questions we split men and women into groups according to their answers to the following question, ``How do you feel about a CS major as of today'' with answers on a 5 point Likert scale (``More likely'', ``Likely'', ``Neutral'', ``Less Likely'', ``Much less likely'')  asked during the Exit Survey for the Spring and Fall 2017 semesters. The groups were: men who are more likely to major in CS, men who are less likely to major in CS, women who are more likely to major in CS, and women who are less likely to major in CS.

\begin{figure}[!htb]
\minipage{0.29\textwidth}
  \includegraphics[width=\linewidth]{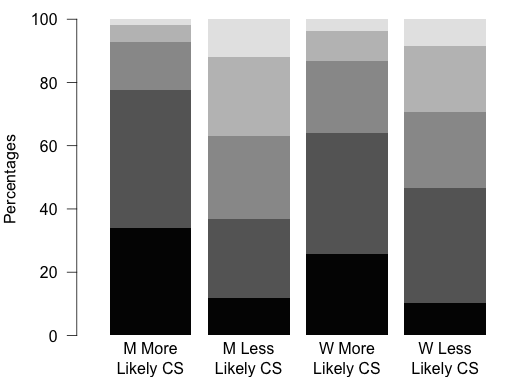}
  \caption{Level of agreement with feeling comfortable asking the \textbf{instructor} questions.}\label{f:instr_mlm}
\endminipage\hfill
\minipage{0.29\textwidth}
  \includegraphics[width=\linewidth]{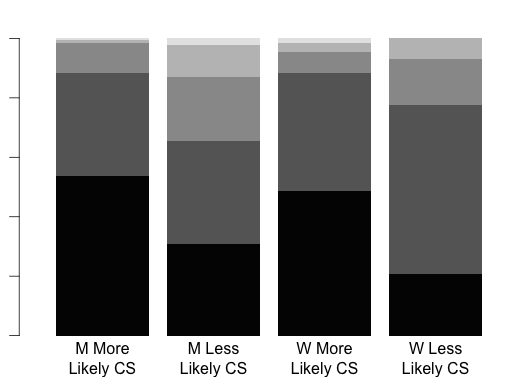}
  \caption{Level of agreement with feeling comfortable asking the \textbf{peer leader} questions.}\label{f:peer_mlm}
\endminipage\hfill
\minipage{0.29\textwidth}%
  \includegraphics[width=\linewidth]{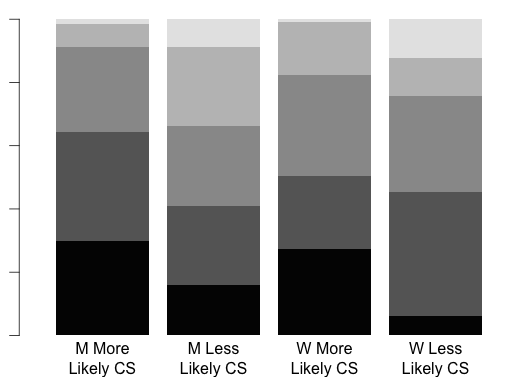}
  \caption{Level of agreement with feeling comfortable asking the \textbf{tutor} questions.}\label{f:tutor_mlm}
\endminipage
\minipage{0.11\textwidth}%
  \includegraphics[width=\linewidth]{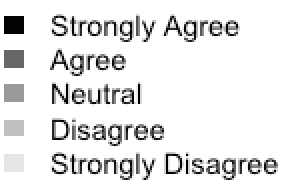}
 \label{f:legend}
\endminipage
\end{figure}

In each group, we computed the distribution of students by level of agreement with statements mentioned above and show the distributions of answers in Figures~\ref{f:instr_mlm},~\ref{f:peer_mlm}, and~\ref{f:tutor_mlm}). We see that students who are more likely to major in CS after taking CS1 are also in stronger agreement with the statement of confidence in asking questions. This result points to a possible opportunity for retention, which we will explore in our future work: if we are able to increase the students' comfort levels in using the various resources offered in the class, would that play into the students' increased interest in and commitment to the major?

\begin{figure}
  \begin{minipage}[c]{0.67\textwidth}
    \includegraphics[width=\textwidth]{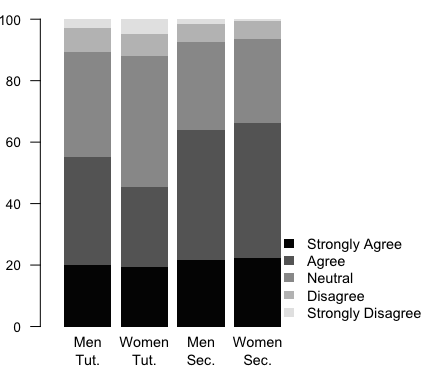}
  \end{minipage}\hfill
  \begin{minipage}[c]{0.3\textwidth}
    \caption{Level of agreement with ``I enjoy interacting with the other students at the [community space]'' for men and women and ``I enjoy interacting with the other students in my section of CS1'' for men and women.}
     \label{f:interaction}
  \end{minipage}
\end{figure}

\subsection{Interaction with Other Students}

Next, we answer the question if men and women differ in how much they enjoy interacting with other students. The question in our Exit survey asks specifically about interaction with other students at the community space and about interaction with other students in the same section of CS1. We show the distribution of answers to the question, \textit{How much do you agree with the following statement: ``I enjoy interacting with the other students at the [community space].'' ?} and \textit{How much do you agree with the following statement, ``I enjoy interacting with other students in my section of CS1''?} in Figure~\ref{f:interaction}.

Within each statement, the difference in level of agreement between men and women is not statistically significant, suggesting that men and women enjoy interacting with other students to a similar extent. However, within the same gender, the difference in level of agreement between the two statements is statistically significant, which leads to the conclusion that both men and women seem to enjoy interacting with other students in their section of CS1 more than they enjoy interacting with other students in the community space. This is important because the community space was designed to encourage interaction between students and should make this interaction enjoyable. The observation that the student interaction in the community space is less enjoyable to students than the student interaction in general points to a need to improve our community space. In our future work, we will explore factors that affect these perceptions.

\subsection{Resources Used}

\begin{figure}[!htb]
\minipage{0.44\textwidth}
  \includegraphics[width=\linewidth]{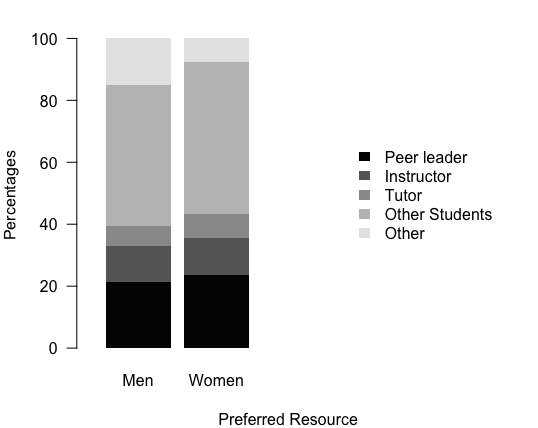}
  \caption{Preferred resources by Gender.}\label{f:res_pref}
\endminipage\hfill
\minipage{0.55\textwidth}%
  \includegraphics[width=\linewidth]{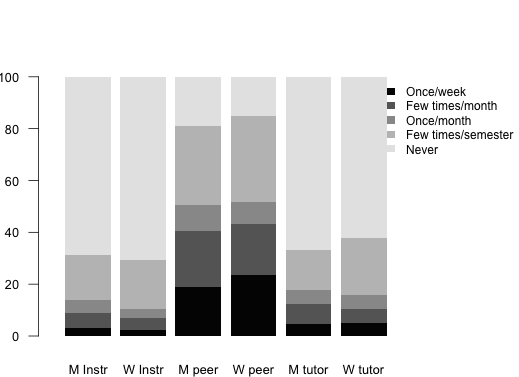}
  \caption{The frequency of using the CS1 resources offered (peer-leader, instructor's office hours, tutor) for men and women.}
  \label{f:res_freq}
\endminipage
\end{figure}

When stuck on their homework or needing help to understand course material, do men and women use the same resources or do they prefer different ones? We use the Exit Survey to answer this question. The Exit Surveys from Spring 2016, Fall 2016, Spring 2017, and Fall 2017 ask students the question: ``If I get stuck on my homework or feel confused about something I learned in class, I prefer to \ldots '' with options, ``Ask the peer leader for help'', ``Ask the instructor for help'', ``Ask the tutors at the [community space]'', ``Ask other students in the class'', ``None of these''. We count how many men and women answered in each of these categories and get the numbers in Figure~\ref{f:res_pref}. The difference between men and women is statistically significant here. According to our data, the resource that students use the most often are other students. A higher proportion of women than men prefer to ask the instructor, tutors and other students for help, whereas a higher proportion of men than women prefer to ask the peer leaders for help.

Having seen what resources students \textit{prefer}, we now turn our attention to the resources students \textit{actually use}, more specifically, we ask if men and women use the same resources with the same frequency? The Exit survey from semesters Spring 2017 and Fall 2017 asked students how often they asked the peer-leader for help, how often they went to the instructor's office hours, and how often they asked the tutors for help. We show the proportion of men and women giving each answer in Figure~\ref{f:res_freq}. We see that both men and women prefer asking the peer-leaders for help at much higher rates than the instructor and the tutors. These numbers point to the success of our peer-leader program in CS1. 

The question about how frequently students used the various resources offered did not include other students, a fact that differentiates the results shown in Figures~\ref{f:res_pref} and \ref{f:res_freq}. However, in Figure~\ref{f:res_pref}, we see that the second most preferred resource after other students in the class is the peer leader, consistently with a more frequent reliance on peer leaders for help (Figure~\ref{f:res_freq}).

\subsection{Summary}

In this section, we have explored environmental factors in our CS1 course such as the gender gap, the students' intent to major in CS, and how much students use the resources offered to them, and explored how they affect men and women in terms of retention, comfort levels, etc. In the next section, we explore factors that have to do with the students' background in computing.

\section{Students' Background}
\label{s:background} 

In this section, we explore differences in the computing background of our female and male students and the possible impact on intent to major and retention.

\subsection{Prior Experience}

\begin{figure}[!htb]
\minipage{0.32\textwidth}
  \includegraphics[width=\linewidth]{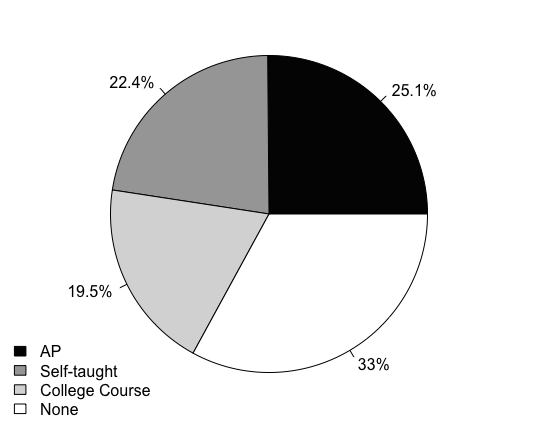}
  \caption{Prior experience for students F12 - S14.}\label{f:prior_f12_s14}
\endminipage\hfill
\minipage{0.32\textwidth}
  \includegraphics[width=\linewidth]{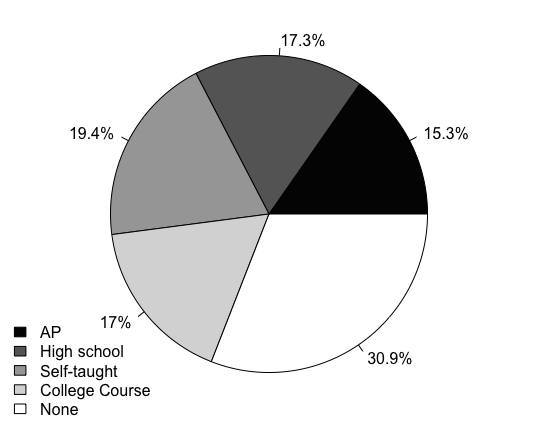}
  \caption{Prior experience for men F15 - F17.}\label{f:prior_men_f15_f17}
\endminipage\hfill
\minipage{0.32\textwidth}%
  \includegraphics[width=\linewidth]{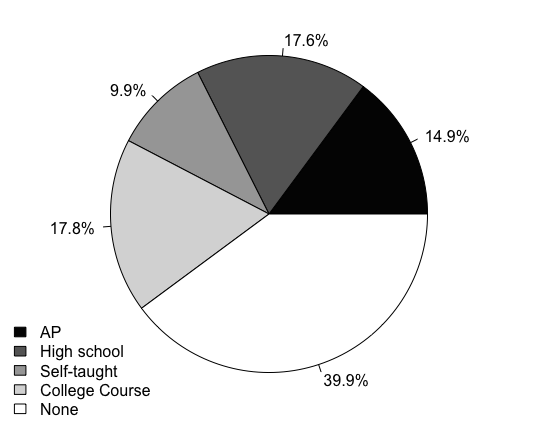}
  \caption{Prior experience for women F15 - F17.}\label{f:prior_women_f15_f17}
\endminipage
\end{figure}

Do men and women who take CS1 have different prior experience in CS? We use the Introductory Survey to answer this question. The surveys completed during Fall 2012, Spring 2013, Fall 2013, and Spring 2014 asked students what was their prior experience in computing. The possible answers were: ``High school AP'', ``Self-taught'', ``Another college course'', and ``None''. When computing the proportion of men and women with these different prior experiences, the differences were not statistically significant. We combined the numbers for men and women and show the proportion of student with these different experiences in Figure~\ref{f:prior_f12_s14}.

Have these proportions changed in recent years? We use data from the Introductory Survey from Spring 2016, Spring 2017, and Fall 2017. The survey administered during these semesters added ``High school'' as another possible answer, when asking students about their prior experience in computing. We show the distribution of prior experience for men and women using data from the three semesters mentioned above in Figure~\ref{f:prior_men_f15_f17} for men and in Figure~\ref{f:prior_women_f15_f17} for women. We notice a recent significant drop in the percentage of women who have taken another college course before CS1 and that more women recently have no prior experience in computing.

\subsection{Continuation Rates By Prior Experience}

Do some kinds of prior experience correlate with higher continuation rates for men and women? Similarly to the previous section, we answer this question by using the students' answers to the Introductory Survey. We use the combined data from Fall 2012, Spring 2013, Fall 2013, and Spring 2014. These are the semesters for which the survey question ``What is your prior experience?'' offered the same possible answers. We determine how many students have each kind of prior experience (AP, self-taught, college course, none), then count how many students in each subgroup take CS2.  We show the numbers in Table~\ref{t:cont_prior_old}.

\begin{table}[t]
\centering
\caption{Continuation rates to CS2 by Prior Experience (Fall 2012, Spring 2013, Fall 2013, Spring 2014).Percentages are calculated for each group (for example, women with AP experience, what proportion continue and what proportion do not continue.}
\begin{tabular}{crrrr} \hline
Prior Experience&Not Cont. Men& Not Cont. Women &Cont. Men & Cont. Women\\ \hline
AP & $67$ ($33.2$\%)&$56$ ($32.7$\%)& $135$ ($66.8$\%) & $115$ ($67.3$\%)\\ \hline
Self-Taught & $62$ ($34.1$\%)& $56$ ($37.1$\%) & $120$ ($65.9$\%) & $95$ ($62.9$\%) \\ \hline
College & $81$ ($50.6$\%) &  $68$ ($52.7$\%) & $79$ ($49.4$\%)& $61$ ($47.3$\%) \\ \hline
 None & $129$ ($48.5$\%)  & $115$ ($51.6$\%) & $137$ ($51.5$\%) & $108$ ($48.4$\%)\\ \hline
\end{tabular}
\label{t:cont_prior_old}
\vspace{-0.2in}
\end{table}

For both men and women, the difference in Table~\ref{t:cont_prior_old} between students who take CS2 and students who do not take CS2 is statistically significant. We conclude that, \emph{for both men and women, students who have taken an AP class or have taught themselves programming are more likely to continue to CS2 than students who have taken a computing related college course or do not have any prior experience.}

For students who continue in CS, the difference between the distributions of prior experiences for men and women are not statistically significant. 

\subsection{Prior Experience and Retention}
\label{ss:prior_exp_retention}

Is there a correlation between majoring in CS and the students' various kinds of prior experience? We use the data set from semesters Fall 2012, Spring 2013, Fall 2013, and Spring 2014, and count how many men and women with various prior experience take CS4 (Table~\ref{t:major_prior_old}), which is a good approximation of the students who are majoring, since most of the students who take CS4 are CS majors .

\begin{table}[t]
\centering
\caption{Continuation rates to CS4 by Prior Experience (Fall 2012, Spring 2013, Fall 2013, Spring 2014).}
\begin{tabular}{crrrr} \hline
Prior Experience&Not major Men& Not major Women &Major Men & Major Women\\ \hline
AP & $114$ ($22.8$\%)&$97$ ($22.8$\%)& $88$ ($28.4\%$) & $74$ ($29.7$\%)\\ \hline
Self-Taught & $99$ ($19.8$\%)& $84$ ($19.7$\%) & $83$ ($26.8$\%) & $67$ ($26.9$\%) \\ \hline
College & $107$ ($21.4$\%) &  $86$ ($20.2$\%) & $53$ ($17.1$\%)& $43$ ($17.3$\%) \\ \hline
None & $180$ ($36.0$\%)  & $158$ ($37.2$\%) & $86$ ($27.7$\%) & $65$ ($26.1$\%)\\ \hline
\end{tabular}
\label{t:major_prior_old}
\vspace{-0.2in}
\end{table}

Among students who end up majoring in CS and among students who end up not majoring in CS, there is no statistically significant difference between men and women in terms of the proportion of students with various prior experiences. However, for both gender groups, \emph{the difference in the distributions of prior experiences of students who major in CS and students who do not major in CS is statistically significant.} Among students who end up majoring in CS, a higher proportion have AP experience ($28.4\%$ vs $22.8\%$ for men and $29.7\%$ vs $22.8\%$ for women) and are self-taught ($26.8\%$ vs $19.8\%$ for men and $26.9\%$ vs $19.7\%$ for women). A higher proportion of students who choose to not major in CS have previously taken a college course or have no prior experience. In the previous section we have seen that students who have AP experience or have taught themselves programming are more likely to take the follow up class CS2. In this section we see that this trend seems to continue to CS4.

\subsection{Gender Difference in Programming Experience}

In this section, we explore whether or not there is a gender difference in the proportion of students who have taken a programming course prior to CS1. We use the Introductory Survey data, more specifically, the students' answers to the question, ``Have you taken any programming courses prior to this course?'', which is a yes/no question. We asked this question during the following semesters, Fall 2015, Spring 2016, Fall 2016, Spring 2017, and Fall 2017. We show the counts of students in each category in Table~\ref{t:prior_prog_c}.

\begin{table}[t]
\centering
\caption{Programming Experience for Men and Women by Semester. ``Men Prog.'' means ``Men with programming experience'', ``Men No Prog.'' means ``Men without programming experience'', ``Women Prog.'' means ``Women with programming experience'', and ``Women No Prog.'' means ``Women without programming experience''. For semesters marked with an ``*'', the difference between the proportions of men and women are statistically significant.}
\begin{tabular}{lrrrr} \hline
Semester&Men Prog.& Men No Prog. & Women Prog. & Women No Prog.\\ \hline
Fall 2015$^*$ & $248$ ($55.5$\%)&$199$ ($44.5$\%)& $66$ ($44.0$\%) & $84$ ($56.0$\%)\\ \hline
Spring 2016 & $140$ ($46.7$\%)& $160$ ($53.3$\%) & $48$ ($41.7$\%) & $67$ ($58.3$\%) \\ \hline
Fall 2016$^*$ & $292$ ($56.6$\%) &  $224$ ($43.4$\%) & $74$ ($40.9$\%)& $107$ ($59.1$\%) \\ \hline
Spring 2017 & $85$ ($48.9$\%)  & $89$ ($51.1$\%) & $49$ ($49.5$\%) & $50$ ($50.5$\%)\\ \hline
Fall 2017 & $207$ ($55.6$\%)  & $165$ ($44.4$\%) & $88$ ($46.6$\%) & $101$ ($53.4$\%)\\ \hline
\end{tabular}
\label{t:prior_prog_c}
\vspace{-0.2in}
\end{table}

In general, for students who have previously taken a programming course, the proportion of men is higher than the proportion of women, even though the difference in the proportion of men and women is only statistically significant for the Fall 2015 and the Fall 2016 semesters. Existing literature also reports that men are more likely than women to have prior experience in computing~\cite{margolis03}.

\subsection{Encouragement to Pursue CS}

Previous work \cite{margolis03}  discussed how men and women received different levels of encouragement to pursue CS because of the stereotype that CS is a masculine field. Related work \cite{cohoon01} speculates that one reason the computer science field retains women at much lower rates than men is because men receive more encouragement to pursue CS than women.

We were interested to see if our data confirmed these previous observations. Our Introductory Survey asked students if they received encouragement to pursue CS. We used our data from semesters Fall 2015, Spring 2016, Spring 2017, and Fall 2017. We counted how many men and women said they received encouragement to pursue CS and how many said they did not receive encouragement to pursue the major. Encouragingly, the difference in the proportion of students who received encouragement between men and women (Table~\ref{t:encouragement}) is not statistically significant.

\begin{table}[t]
\centering
\caption{Men and Women in CS1 who have or have not received encouragement to pursue CS.}
\begin{tabular}{ccc} \hline
&Received Encouragement & Did Not Receive Encouragement \\ \hline
Men & $900$ ($70.6$\%)&$375$ ($29.4$\%)\\ \hline
Women & $409$ ($74.9$\%)& $137$ ($25.1$\%) \\ \hline
\hline\end{tabular}
\label{t:encouragement}
\vspace{-0.2in}
\end{table}

\begin{figure}
\centering
\includegraphics[width=.6\textwidth]{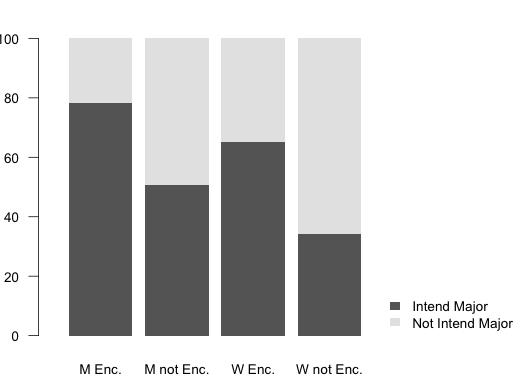}
\caption{Difference in the proportion of students who intend to major between students who received encouragement to pursue CS and students who did not receive encouragement to pursue CS. For both Men and Women, the difference is statistically significant.}
\label{f:enc_i_major}
\end{figure}

\begin{figure}[htp]
\hspace*{\fill}%
\begin{minipage}[t]{0.48\textwidth}
\centering
\vspace{0pt}
  \includegraphics[width=\linewidth]{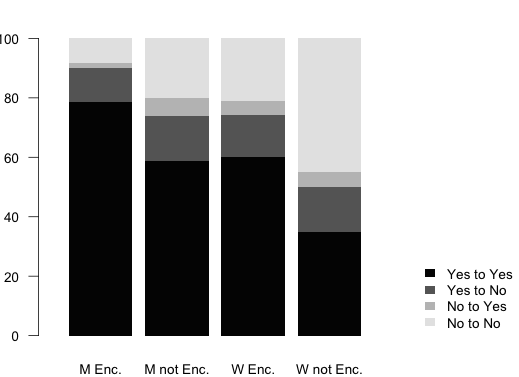}
  \caption{Difference in the proportion of students who changed or did not change their minds about majoring in CS after taking CS1 between students who received encouragement to pursue CS and students who did not receive encouragement to pursue CS. For both Men and Women, the difference between the two groups is statistically significant.}
  \label{f:enc_ch_maj}
\end{minipage}
\hfill
\begin{minipage}[t]{0.48\textwidth}%
\centering
\vspace{0pt}
  \includegraphics[width=\linewidth]{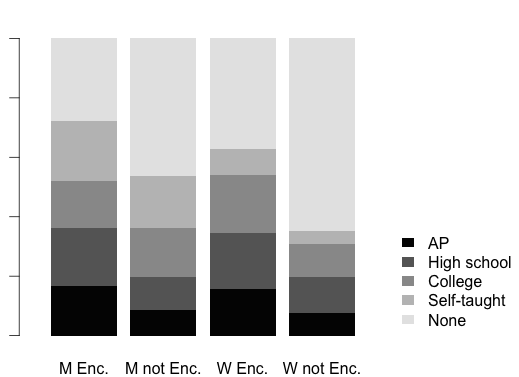}
  \caption{Difference in prior experience between students who were encouraged and students who were not encouraged to pursue CS.}
  \label{f:enc_prior}
\end{minipage}
\end{figure}

Figures~\ref{f:enc_i_major}, \ref{f:enc_ch_maj}, and \ref{f:enc_prior} show the relationship between having received encouragement to pursue CS and intent to major, prior experience, and a change in intent to major in CS after having taken CS1, respectively. The data used to generate the first and the third graph was the Registrar data and Introductory Survey data from semesters Fall 2015, Spring 2016, Spring 2017, and Fall 2017. The data used to generate the second graph was the Registrar and the Introductory Survey data from semesters Spring 2016, Fall 2016, and Spring 2017.
Prior experience is important to consider in our analysis because it is linked to a higher likelihood of retention (as we showed in Section~\ref{ss:prior_exp_retention}).

In Figure~\ref{f:enc_i_major}, we see, for both men and women, that students who received encouragement to pursue CS being more likely to intend to major in CS. In Figure~\ref{f:enc_ch_maj}, for both gender groups, students who have received encouragement to pursue CS are more likely to be in the ``yes to yes'' category (that is, having intended to major at the beginning of CS1 and intending to major at the end of CS1). In Figure~\ref{f:enc_prior}, we see that the proportion of men and women who have taken an AP class is about the same within the group of students who have received encouragement to pursue CS and within the group of students who have not received encouragement to pursue CS. 

\subsection{Gender Difference in Self-perceived Interests and Abilities}

Previous research pointed to gender differences in computer self-efficacy~\cite{beyer14} and to a relationship between the performance of women in CS and their self-assessment of technical abilities~\cite{cohoon01}. Others have suggested a connection between the retention of women in computing and their self-confidence~\cite{shapiro11}. In the following, we explore gender differences in the CS1 students' self-assessed computing related abilities and interests. 

\begin{figure}
\centering
\includegraphics[width=.9\textwidth]{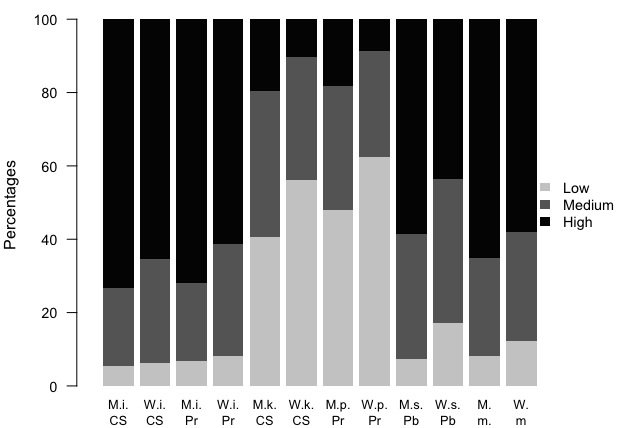}
\caption{Men and women rated their perceived Interest in CS (M.i.CS and W.i.CS respectively), Interest in Programming (M.i.Pr and W.i.Pr), CS knowledge (M.k.CS and W.k.CS), Programming proficiency (M.p.Pr and W.p.Pr), Problem Solving ability (M.s.PB and W.s.Pb), and Math ability (M.m and W.m).}
\label{f:interests_prof}
\end{figure}

In the Introductory Survey, we asked our students to rate themselves in the following categories: their interest in CS, their interest in programming, their CS knowledge, their programming proficiency, their problem solving ability, and their math ability. We used data from semesters Spring 2017 and Fall 2017 to measure these perceptions and show the results in Figure~\ref{f:interests_prof}. The difference between men and women is statistically significant for \emph{interest in programming}, \emph{CS knowledge}, \emph{programming proficiency}, and \emph{problem solving ability}.  The higher self assessment of men compared to women is consistent with previous results \cite{margolis03, cohoon01}.

\subsection{Correlation between Knowing Someone in CS and Intent to Major}

Existing research discusses the importance of encouragement to study CS in attitudes toward the major, especially for women~\cite{margolis03}. We asked the question, for each gender group, if knowing someone in the field correlates with a higher likelihood of intending to major in CS. To answer this question, we use Introductory Survey data from Fall 2016. This survey asked students if they knew someone in CS and who that person was. We split our students into men and women and, for each group, split the students again by whether or not they intended to major and by whether or not they knew someone in CS. We show these numbers in Table~\ref{t:know_someone}.

\begin{table}[!htbp]
\centering
\caption{Intent to Major by Knowing Someone in CS.}
\begin{tabular}{*5c}
\toprule
Know Somone in CS? &  \multicolumn{2}{c}{Men} & \multicolumn{2}{c}{Women}\\
\midrule
{}   & Intend Major   & Not Intend Major    & Intend Major   & Not Intend Major\\
Yes   &  $357$ ($81.5\%$) & $81$ ($18.5\%$)   & $110$ ($70.1\%$)  & $47$ ($29.9\%$)\\
No  & $59$ ($75.6\%$) & $19$ ($24.4\%$)  & $19$ ($79.2\%$)  & $5$ ($20.8$)\\
\bottomrule
\end{tabular}
\label{t:know_someone}
\end{table}

For both men and women, the difference in the proportion of students who intend to major in CS between those who know someone in CS and those who do not is not statistically significant. However, if we look within the group of students who know someone in CS, the difference in the proportion of students who intend to major in CS between men and women is statistically significant. For students who do not know someone in CS, this difference is not statistically significant. We conclude that our data does not provide enough evidence that knowing someone in CS is correlated in a significant way to intent to major in CS.

\subsection{Gender Difference in the Use of Technology}

Finally, our Introductory Survey for semesters Fall 2015, Spring 2016, and Spring 2017 asks students how many hours per week they spent, on average, using the Internet, playing video games, and using productivity software. We are interested in finding out if there are significant differences between men and women in these areas because existing literature reports a higher percentage of men are  comfortable with technology and, therefore, feel more at home in a technical field such as computer science \cite{margolis03}.

We combined the data on these questions for the three semesters mentioned above, and, for each category (Internet, games, and productivity software) ran a t-test to determine the significance of the difference between men and women in the number of hours reported. The only category where the difference is significant is video games, with an average of about $11$ hours per week for men, and $5.7$ hours per week for women. Men and women say they use the Internet for an average of about $29$ hours per week, and productivity software for an average of about $16$ hours per week.

\subsection{Summary}

In this section, we have explored gender differences in the CS1 students' computing background and have seen that fewer women than men have prior experience in computing and that more men than women have taken a CS AP class or have taught themselves programming. We have also seen that students with AP experience and those who have taught themselves programming are more likely to take the follow-up CS2 class and that a higher percentage of students who end up majoring in CS have AP experience or have taught themselves programming. We have also not found evidence that men and women receive different levels of encouragement to pursue CS, but that students who have received this encouragement are more likely to intend to major both before and after taking CS1. Other than a significant difference in the use of video games, with men spending more time playing video games than women, our data does not support a significant difference between men and women in their use of technology.

\section{Students' Grades}
\label{s:grades}

In this section, we focus on the students' grades in the CS1 class and how they correlate with various other factors, such as intent to major in CS and continuation rates. 

\subsection{Grades and Continuation to CS2}

\begin{figure}
\centering
\includegraphics[width=0.7\textwidth]{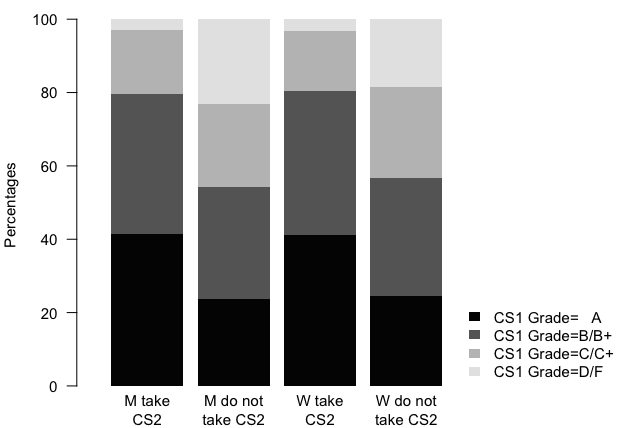}
\caption{Grade distribution for men (M) who take CS2, men (M) who do not take CS2, women (W) who take CS2, and women (W) who do not take CS2}
\label{f:grade_dist_by_contMW}
\end{figure}

We start by exploring the correlation between grades and the decision to take CS2 for men and women. We split our students by gender and by whether or not they take CS2. We show the grade distribution for each group in Figure~\ref{f:grade_dist_by_contMW}.

We see that, for both men and women, the proportion of students who get high grades (As, Bs and B+s) is much higher for students who end up taking CS2, whereas the proportion of students who get low grades is much higher for students do not take CS2. It is interesting to note that \emph{the difference in grade distribution between men and women who continue to CS2 is not statistically significant}. For students who do not take CS2, women get slightly higher grades than men, and this difference is statistically significant.  This result is consistent with a previously reported correlation between retention and self-efficacy~\cite{beyer14}, and grades are significant contributors to the students' sense of self-efficacy~\cite{wood87}.

\subsection{Gender Difference in Continuation Rates}

Next, we ask whether there is a difference in continuation rates between men and women, controlling for intent to major before taking CS1 and grades. We show the student counts in each group in Table \ref{t:grades_cont_int_major} for students who intend to major in CS at the beginning of CS1 and in Table~\ref{t:grades_cont_not_int_major} for students who do not intend to major in CS at the beginning of CS1. These numbers are generated using the Registrar data set for semesters Fall 2012 though Spring 2017.

\begin{table}
\centering
\caption{Continuation rates by Gender by CS1 Grades for Students who Intend to Major in CS.}
\subcaption*{CS1 Grade = A}
\begin{tabular}{crr} \hline
&Do not take CS2& Do take CS2\\ \hline 
Female & $75$ ($36.9$\%) & $128$ ($63.1$\%)\\ \hline
Male & $235$ ($31.5$\%) & $511$ ($68.5$\%)\\ \hline
\end{tabular}
\bigskip
\subcaption*{CS1 Grade = B/B+}
\begin{tabular}{crr} \hline
&Do not take CS2&Do take CS2\\ \hline  
Female & $81$ ($43.8$\%) & $104$ ($56.2$\%)\\ \hline
Male & $205$ ($31.9$\%) & $438$ ($68.1$\%)\\ \hline
\end{tabular}
\bigskip
\subcaption*{CS1 Grade = C/C+}
\begin{tabular}{crr} \hline
&Do not take CS2&Do take CS2\\ \hline  
 Female & $64$ ($61.0$\%) & $41$ ($39.0$\%)\\ \hline
Male & $172$ ($50.1$\%) & $171$ ($49.9$\%)\\ \hline
\end{tabular}
\bigskip
\subcaption*{CS1 Grade = D/F}
\begin{tabular}{crr} \hline
&Do not take CS2&Do take CS2\\ \hline  
Female & $55$ ($88.7$\%) & $7$ ($11.3$\%)\\ \hline
Male & $173$ ($88.7$\%) & $22$ ($11.3$\%)\\ \hline
\end{tabular}
\label{t:grades_cont_int_major}
\end{table}

\begin{table}
\centering
\caption{Continuation rates by Gender by CS1 Grades for Students who do not Intend to Major in CS.}
\subcaption*{CS1 Grade = A}
\begin{tabular}{crr} \hline
&Do not take CS2&Do take CS2\\ \hline 
Female & $85$ ($66.9$\%) & $42$ ($33.1$\%)\\ \hline
Male & $103$ ($55.1$\%) & $84$ ($44.9$\%)\\ \hline
\end{tabular}
\bigskip
\subcaption*{CS1 Grade = B/B+}
\begin{tabular}{crr} \hline
&Do not take CS2&Do take CS2\\ \hline  
Female & $104$ ($80.0$\%) & $26$ ($20.0$\%)\\ \hline
Male & $144$ ($73.8$\%) & $51$ ($26.2$\%)\\ \hline
\end{tabular}
\bigskip
\subcaption*{CS1 Grade = C/C+}
\begin{tabular}{crr} \hline
&Do not take CS2&Do take CS2\\ \hline  
Female & $69$ ($86.2$\%) & $11$ ($13.8$\%)\\ \hline
Male & $97$ ($78.2$\%) & $27$ ($21.8$\%)\\ \hline
\end{tabular}
\bigskip
\subcaption*{CS1 Grade = D/F}
\begin{tabular}{crr} \hline
&Do not take CS2&Do take CS2\\ \hline  
Female & $41$ ($91.1$\%) & $4$ ($8.9$\%)\\ \hline
Male & $82$ ($95.3$\%) & $4$ ($4.7$\%)\\ \hline
\end{tabular}
\label{t:grades_cont_not_int_major}
\end{table}

For students who intend to major before taking CS1 and end up getting a B or a B+ in the class, and for students who do not intend to major in CS before taking CS1 and receive an A in the class, women are less likely to take CS2 than men. For all other groups, the difference in continuation rates between men and women is not statistically significant. As we gather more data in the future, we believe that some of these differences may also become statistically significant.

\subsection{Difference in Grades between Students who Changed their Mind about Majoring and Students who did not Change their Mind about Majoring}

For men and women, is there a difference in grades between those who change their minds about majoring in CS after taking CS1 and those who do not? To answer this question, we use the Registrar and the Exit survey data sets. The Exit survey asks students how likely they were to major in CS before taking CS1 and how likely they are to major after taking CS1. By combining answers to these two questions, we can determine which students changed their intent to major in CS and which students did not. We split our students in two groups by gender, and by whether or not they changed their intent to major after taking CS1. Then, we count how many students got As, Bs, Cs, and Ds and Fs in each group and run a chi-square test for each gender group to determine if the grades are different for students who changed their intent to major compared to students who did not change their intent to major. We show these numbers in Table~\ref{t:change_no_change}. The number of students who went from not intending to major to intending to major were too small to compute statistical significance, and they are not shown here.

\begin{table}
\centering
\caption{Difference in Grades between Students who Do not Change their Minds about Majoring and Students who Change their Minds from Majoring to Not Majoring after Taking CS1. The percentages for men are computed using the total number of male students who have taken the exit survey (Fall 15 - Fall 17) and, for women, using the total number of female students who have taken the exit survey.}
\subcaption*{Men}
\begin{tabular}{ccccc} \hline
Grade& A& B& C& D and F\\ \hline 
Majoring to Majoring & $464$ ($44.2$\%) & $370$ ($35.2$\%)& $154$ ($14.7$\%) & $62$ ($5.9$\%)\\ \hline
Majoring to Not Majoring  & $23$ ($18.9$\%) & $34$ ($27.9$\%)& $29$ ($23.8$\%) & $36$ ($29.5$\%)\\ \hline
\end{tabular}
\bigskip
\subcaption*{Women}
\begin{tabular}{ccccc} \hline
Grade&A& B& C& D and F\\ \hline 
Majoring to Majoring & $138$ ($45.8$\%) & $111$ ($36.9$\%)& $36$ ($12.0$\%) & $16$ ($5.3$\%)\\ \hline
Majoring to Not Majoring  & $7$ ($13.2$\%) & $14$ ($26.4$\%)& $18$ ($34.0$\%) & $14$ ($26.4$\%)\\ \hline
\end{tabular}
\label{t:change_no_change}
\end{table}

For both men and women, students who did not change their minds about majoring after taking CS1 received, overall higher grades than students who did change their minds about majoring ($p<0.05$). Also, for students who received a C in CS1, there is a statistically significant difference between men and women in the proportion of students who went from intending to major before taking the class to not intending to major after taking the class.

\subsection{Correlation between Early Grades and Later Grades}

To determine whether or not grades in courses taken earlier in the CS major correlate with grades in courses taken later in the CS major, we used the Registrar Data set, more specifically, information about the grades that students received in CS1, CS2, CS3, and CS4 and the students' gender. Since the Registrar data contains letter grades, we used the Goodman and Kruskal's G test \cite{goodman54} to assess the strength of the association between grades in every pair of courses.

We find that there is a \emph{significant moderate positive association between every pair of courses}. The values of the Goodman and Kruskal's G for all students, men, and women, are shown in Table \ref{t:grade_pred_early_later}. 

\begin{table}[h]
\centering
\caption{Goodman and Kruskal's G Coefficient for Earlier and Later Grades for all students, only male students, and only female students.}
\begin{tabular}{crrr} \hline
&All& Men& Women\\ \hline 
 CS1 vs. CS2 & $0.475$  & $0.464$  & $0.517$\\ \hline
 CS1 vs. CS3 & $0.400$  & $0.414$ & $0.323$\\ \hline
 CS1 vs. CS4 & $0.334$  & $0.310$  & $ 0.442$\\ \hline
 CS2 vs. CS3 & $0.405$  & $0.400$ & $0.422$\\ \hline
 CS2 vs. CS4 & $0.377$  & $0.381$  & $0.365$\\ \hline
 CS3 vs. CS4 & $0.415$  & $0.427$ & $0.364$\\ \hline
\end{tabular}
\label{t:grade_pred_early_later}
\end{table}

We do not have insight into the why of this correlation, but these results do intuitively suggest the need to work hard to students truly understand CS material and earn good grades early in their education.

\subsection{Difference in Grades between Men and Women with the Same Prior Experience}

To determine if there is a difference in grades between men and women with the same prior experience, we use the Registrar data and the Introductory Survey data. The Introductory Survey asked students about their prior experience or CS knowledge. For various semesters, the options students were given to answer this questions varied. We chose to use the data from three semesters (Spring 2016, Spring 2017, and Fall 2017) ($n=1247$) where the options given were identical: AP CS, some high school classes, college, self-learned, and none. We split our students by prior experience and, in each group, used a chi-square test to determine whether or not there was a statistically significant difference in grades between men and women. We show these numbers in Table \ref{t:grades_experience}. 

\begin{table}
\centering
\caption{CS1 Grades by Gender by Prior Experience}
\subcaption*{AP Experience}
\begin{tabular}{crr} \hline
Grade & Men & Women\\ \hline  
A & $67$ ($63.8$\%) & $30$ ($58.8$\%)\\ \hline
B/B+ & $26$ ($24.8$\%) & $13$ ($25.5$\%)\\ \hline
C/C+ & $11$ ($10.5$\%) & $4$ ($7.84$\%)\\ \hline
D/F & $1$ ($0.9$\%) & $4$ ($7.8$\%)\\ \hline
\end{tabular}
\bigskip
\subcaption*{High school Experience}
\begin{tabular}{crr} \hline
Grade & Men& Women\\ \hline  
 A & $41$ ($33.1$\%) & $31$ ($50.0$\%)\\ \hline
 B/B+ & $49$ ($39.5$\%) & $20$ ($32.3$\%)\\ \hline
 C/C+ & $19$ ($15.3$\%) & $8$ ($12.9$\%)\\ \hline
 D/F & $15$ ($12.1$\%) & $3$ ($48.4$\%)\\ \hline
\end{tabular}
\bigskip
\subcaption*{College Course Experience}
\begin{tabular}{crr} \hline
Grade & Men & Women\\ \hline  
 A & $50$ ($41.7$\%) & $27$ ($42.2$\%)\\ \hline
 B/B+ & $35$ ($29.2$\%) & $20$ ($31.2$\%)\\ \hline
 C/C+ & $23$ ($19.2$\%) & $11$ ($17.2$\%)\\ \hline
 D/F & $12$ ($10.0$\%) & $6$ ($9.4$\%)\\ \hline
\end{tabular}
\bigskip
\subcaption*{Self Taught Experience}
\begin{tabular}{crr} \hline
Grade & Men& Women\\ \hline  
  A & $85$ ($59.4$\%) & $13$ ($44.8$\%)\\ \hline
  B/B+ & $22$ ($15.4$\%) & $11$ ($37.9$\%)\\ \hline
  C/C+ & $24$ ($16.8$\%) & $3$ ($10.3$\%)\\ \hline
  D/F & $12$ ($8.4$\%) & $2$ ($6.9$\%)\\ \hline
\end{tabular}
\bigskip
\subcaption*{No Prior Experience}
\begin{tabular}{crr} \hline
Grade & Men& Women\\ \hline  
  A & $107$ ($49.3$\%) & $49$ ($30.6$\%)\\ \hline
  B/B+ & $69$ ($31.8$\%) & $56$ ($35.0$\%)\\ \hline
  C/C+ & $44$ ($5.1$\%) & $34$ ($21.2$\%)\\ \hline
  D/F & $30$ ($13.8$\%) & $21$ ($13.1$\%)\\ \hline
\end{tabular}
\label{t:grades_experience}
\end{table}

Only for students with no prior experience in computing was the difference in the distribution of final grades between men and women statistically significant.

It is worthwhile to test the statistical significance of differences in grade distributions between categories of students with similar prior experience. It points to the importance of prior experience and its relationship to student retention (see Section \ref{ss:prior_exp_retention}).

\subsection{Familiarity with Java and Grades}

Our CS1 class teaches programming in Java. We were interested to explore if familiarity with Java before the start of the class correlated with the students' final grades for the class.

\begin{figure}
\centering
\includegraphics[width=0.7\textwidth]{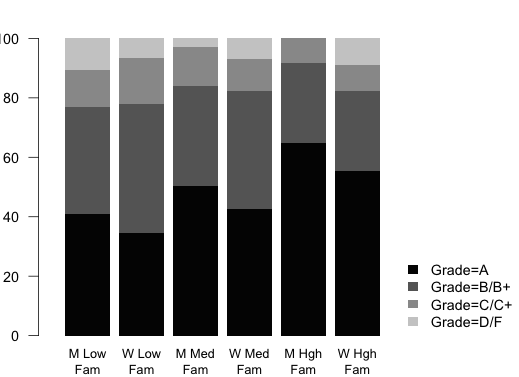}
\caption{Grade distribution for men and women who rate themselves low on the scale of familiarity with Java (``W Low Fam'' and ``M Low Fam''), men and women who rate themselves as having a medium familiarity with Java (``M Med Fam'' and ``W Med Fam'') , and men and women who rate themselves as having a high familiarity with Java (``M High Fam'' and ``W High Fam'').}
\label{f:fam_java}
\end{figure}

We used the Registrar and Introductory Survey data and split students by gender and their self assessed familiarity with Java, an information we gathered during o Spring 2017 and Fall 2017. We split their answers into low, medium, and high familiarity with Java. For each group, we determined the distribution of grades and show these distributions in Figure~\ref{f:fam_java}. 

Within the group of students who rate themselves in the same class of familiarity with Java (low, medium, or high), the only group with a statistically significant difference between men and women is the group of students who rated themselves as \emph{highly familiar} with Java.

For \emph{men}, the difference in the distribution of grades among the three groups (low, medium, and high familiarity with Java) is \emph{statistically significant}. For \emph{women}, the difference in the distribution of grades among the three groups is \emph{not statistically significant}. We conclude that experience with Java may be a helpful but not very important factor affecting the success of students in CS.

\subsection{Correlation Between Self-perceived Interests/Abilities and Grades}

\begin{figure}[!htb]
\minipage{0.48\textwidth}
  \includegraphics[width=\linewidth]{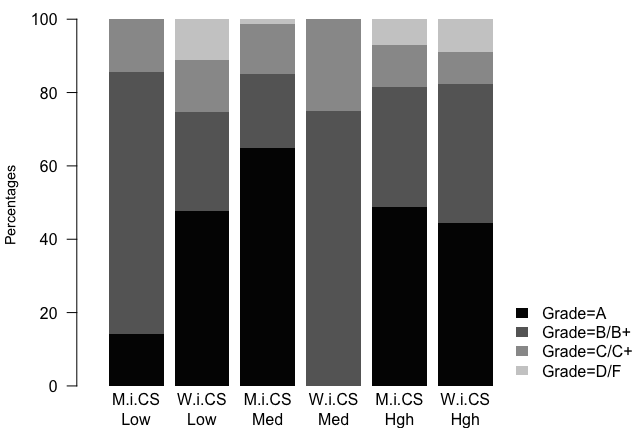}
  \caption{Interest in CS and Grades.}
  \label{f:int_cs_grades}
\endminipage\hfill
\minipage{0.48\textwidth}
  \includegraphics[width=\linewidth]{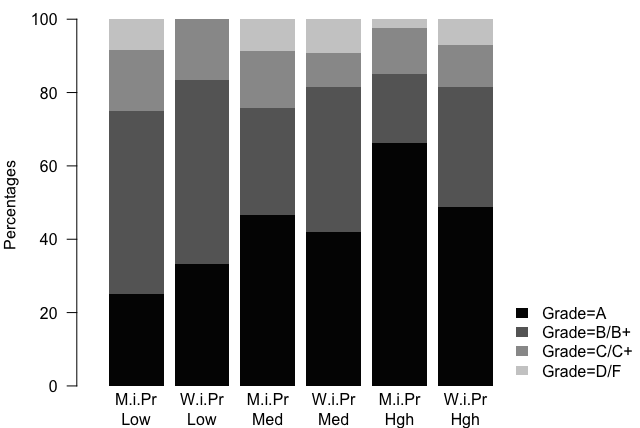}
  \caption{Interest in Programming and Grades.}\label{f:int_prog_grades}
\endminipage
\end{figure}

\begin{figure}[htb]
\minipage{0.48\textwidth}
  \includegraphics[width=\linewidth]{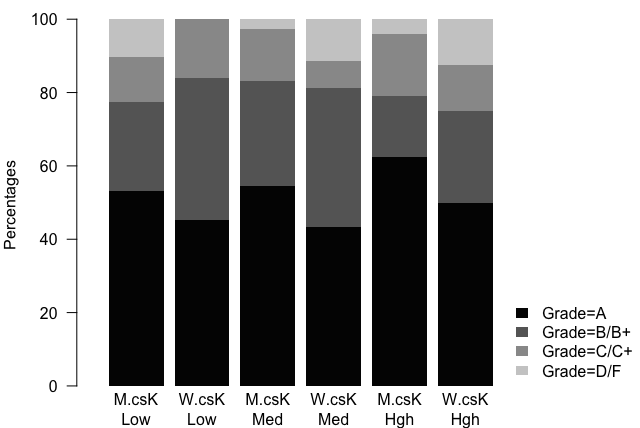}
  \caption{CS Knowledge and Grades.}
  \label{f:CS_know_grades}
\endminipage\hfill
\minipage{0.48\textwidth}
  \includegraphics[width=\linewidth]{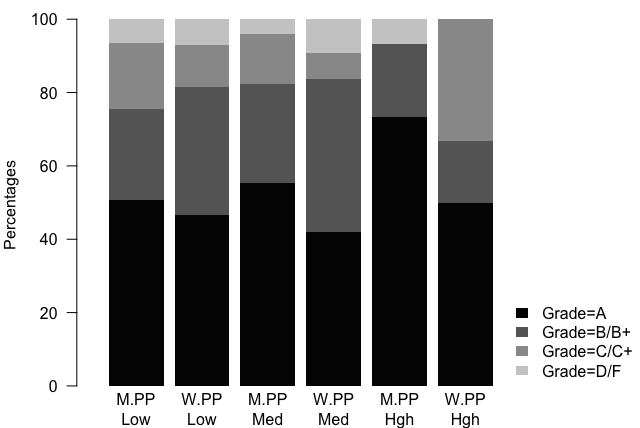}
  \caption{Programming Proficiency and Grades.}
  \label{f:prog_prof_grades}
\endminipage
\end{figure}

\begin{figure}[htb]
\minipage{0.48\textwidth}
  \includegraphics[width=\linewidth]{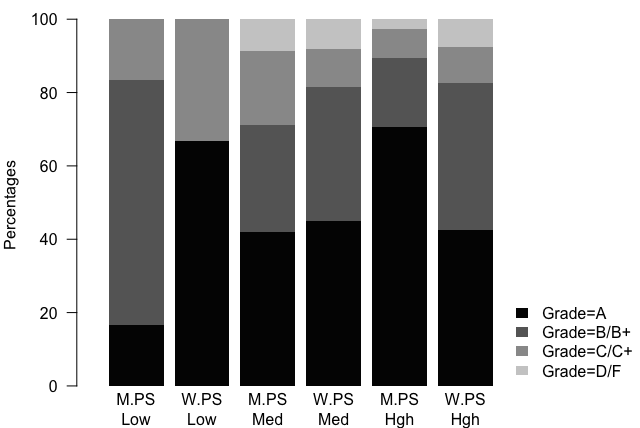}
  \caption{Problem Solving and Grades.}
  \label{f:problem_solving_grades}
\endminipage\hfill
\minipage{0.48\textwidth}
  \includegraphics[width=\linewidth]{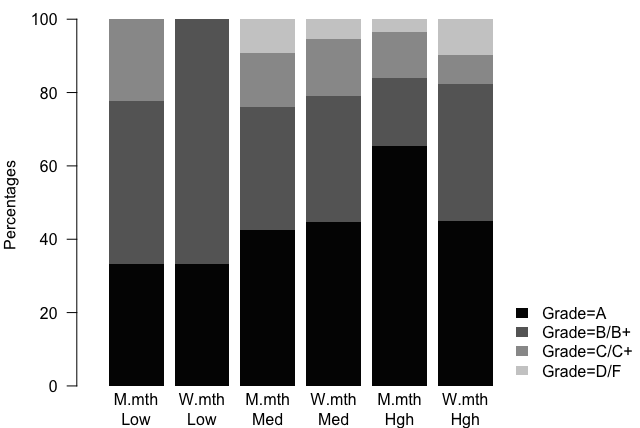}
  \caption{Math Ability and Grades.}
  \label{f:math_grades}
\endminipage
\end{figure}

Figures~\ref{f:int_cs_grades}, \ref{f:int_prog_grades}, \ref{f:CS_know_grades}, \ref{f:prog_prof_grades}, \ref{f:problem_solving_grades}, and \ref{f:math_grades} show the correlation between the students' self assessed interest in CS, interest in programming, CS knowledge, programming proficiency, problem solving ability, math ability, and their grades in CS1, with darker shades of gray signifying higher grades.

In all these six areas, the difference in the proportion of students who received the same grade between men who rated themselves with different abilities and interests is statistically significant, a higher interest or ability correlating with higher grades. Also, the difference between the proportion of students who received the same grade between men and women who rated themselves high in interest in CS, problem solving ability, and math ability is statistically significant, with men in these categories receiving higher grades than women in these categories.

\subsection{Summary}

In this section we have explored the relationship between the students' grades in the CS1 course and other factors such as continuation rates to CS2, a change in intent to major after taking the CS1 course, prior experience in computing, familiarity with Java, and self-perceived interests and abilities.  We have seen that students who take CS2 have higher grades than students who do not, that women are less likely than men to take CS2 even controlling for grades and intent to major, that students who intend to major both before and after taking CS1 have higher grades than students who decide they do not want to major in CS after taking CS1, that, for students with no experience, men are more likely to receive higher grades than women, that experience with the Java programming language may be helpful but not a very important factor in the success of students in CS, and that a high self-assessment of computing-related interests and abilities correlates with higher grades.

\section{Conclusions}

In this paper we have used student data from the CS department at a large public research university to answer questions about men and women taking core courses in the computer science major in three research areas: the CS environment, such as the gender gap in computing, where in the program we lose or fail to attract most of the women, intention to major, comfort levels using various resources offered in CS1; the students' background, such as their prior experience, use of technology, having received encouragement to pursue a career in computing, etc, and how these correlate with retention; and the students' grades and how they relate to factors such as  the decision to take CS2 and familiarity with Java.

It is our hope that other CS departments can use our data and/or methodologies to launch and publish  initiatives to attract and retain more women, and to efficiently and effectively implement changes that will have the greatest positive impact on diversity.

\bibliographystyle{ACM-Reference-Format}
\bibliography{mbabes}


\begin{thebibliography}{38}


\ifx \showCODEN    \undefined \def \showCODEN     #1{\unskip}     \fi
\ifx \showDOI      \undefined \def \showDOI       #1{#1}\fi
\ifx \showISBNx    \undefined \def \showISBNx     #1{\unskip}     \fi
\ifx \showISBNxiii \undefined \def \showISBNxiii  #1{\unskip}     \fi
\ifx \showISSN     \undefined \def \showISSN      #1{\unskip}     \fi
\ifx \showLCCN     \undefined \def \showLCCN      #1{\unskip}     \fi
\ifx \shownote     \undefined \def \shownote      #1{#1}          \fi
\ifx \showarticletitle \undefined \def \showarticletitle #1{#1}   \fi
\ifx \showURL      \undefined \def \showURL       {\relax}        \fi
\providecommand\bibfield[2]{#2}
\providecommand\bibinfo[2]{#2}
\providecommand\natexlab[1]{#1}
\providecommand\showeprint[2][]{arXiv:#2}

\bibitem[\protect\citeauthoryear{Alvarado, Cao, and Minnes}{Alvarado
  et~al\mbox{.}}{2017}]%
        {alvarado17}
\bibfield{author}{\bibinfo{person}{Christine Alvarado},
  \bibinfo{person}{Yingjun Cao}, {and} \bibinfo{person}{Mia Minnes}.}
  \bibinfo{year}{2017}\natexlab{}.
\newblock \showarticletitle{Gender Differences in Students' Behaviors in CS
  Classes throughout the CS Major}. In \bibinfo{booktitle}{\emph{Proceedings of
  the 2017 ACM SIGCSE Technical Symposium on Computer Science Education}}. ACM,
  \bibinfo{pages}{27--32}.
\newblock


\bibitem[\protect\citeauthoryear{Alvarado and Dodds}{Alvarado and
  Dodds}{2010}]%
        {alvarado10}
\bibfield{author}{\bibinfo{person}{C. Alvarado} {and} \bibinfo{person}{Z.
  Dodds}.} \bibinfo{year}{2010}\natexlab{}.
\newblock \showarticletitle{Women in {C}{S}: An {E}valuation of {T}hree
  {P}romising {P}ractices}. In \bibinfo{booktitle}{\emph{Proceedings of the
  SIGCSE Conference}}.
\newblock
\showISBNx{978-1-4503-0006-3}
\urldef\tempurl%
\url{https://doi.org/10.1145/1734263.1734281}
\showDOI{\tempurl}


\bibitem[\protect\citeauthoryear{Babe\c{s}-Vroman, Juniewicz, Lucarelli, Fox,
  Nguyen, Tjang, Haldeman, Mehta, and Chokshi}{Babe\c{s}-Vroman
  et~al\mbox{.}}{2017}]%
        {babes17}
\bibfield{author}{\bibinfo{person}{M. Babe\c{s}-Vroman}, \bibinfo{person}{I.
  Juniewicz}, \bibinfo{person}{B. Lucarelli}, \bibinfo{person}{N. Fox},
  \bibinfo{person}{T. Nguyen}, \bibinfo{person}{A. Tjang}, \bibinfo{person}{G.
  Haldeman}, \bibinfo{person}{A. Mehta}, {and} \bibinfo{person}{R. Chokshi}.}
  \bibinfo{year}{2017}\natexlab{}.
\newblock \showarticletitle{Exploring Gender Diversity in CS at a Large Public
  R1 Research University}. In \bibinfo{booktitle}{\emph{Proceedings of the 2017
  SIGCSE}}.
\newblock


\bibitem[\protect\citeauthoryear{Beyer}{Beyer}{2014}]%
        {beyer14}
\bibfield{author}{\bibinfo{person}{S. Beyer}.} \bibinfo{year}{2014}\natexlab{}.
\newblock \showarticletitle{{Why Are Women Underrepresented in Computer
  Science? Gender Differences in Stereotypes, Self-efficacy, Values, and
  Interests and Predictors of Future CS Course-taking and Grades}}.
\newblock \bibinfo{journal}{\emph{Computer Science Education}}
  \bibinfo{volume}{24}, \bibinfo{number}{2-3} (\bibinfo{year}{2014}).
\newblock


\bibitem[\protect\citeauthoryear{Blaney and Stout}{Blaney and Stout}{2017}]%
        {blaney17}
\bibfield{author}{\bibinfo{person}{Jennifer~M Blaney} {and}
  \bibinfo{person}{Jane~G Stout}.} \bibinfo{year}{2017}\natexlab{}.
\newblock \showarticletitle{Examining the relationship between introductory
  computing course experiences, self-efficacy, and belonging among
  first-generation college women}. In \bibinfo{booktitle}{\emph{Proceedings of
  the 2017 ACM SIGCSE Technical Symposium on Computer Science Education}}. ACM,
  \bibinfo{pages}{69--74}.
\newblock


\bibitem[\protect\citeauthoryear{Cohoon}{Cohoon}{2001}]%
        {cohoon01}
\bibfield{author}{\bibinfo{person}{J.M. Cohoon}.}
  \bibinfo{year}{2001}\natexlab{}.
\newblock \showarticletitle{What {C}auses {W}omen to {D}iscontinue {P}ursuing
  the {U}ndergraduate {C}omputer {S}cience {M}ajor at {H}igher {R}ates than
  {M}en: {T}oward {I}mproving {F}emale {R}etention in the {C}omputer {S}cience
  {M}ajor}.
\newblock \bibinfo{journal}{\emph{Commun. ACM}} \bibinfo{volume}{44},
  \bibinfo{number}{5} (\bibinfo{year}{2001}).
\newblock


\bibitem[\protect\citeauthoryear{Cohoon}{Cohoon}{2002}]%
        {cohoon02}
\bibfield{author}{\bibinfo{person}{J.M. Cohoon}.}
  \bibinfo{year}{2002}\natexlab{}.
\newblock \showarticletitle{Recruiting and {R}etaining {W}omen in
  {U}ndergraduate {C}omputing {M}ajors}.
\newblock \bibinfo{journal}{\emph{ACM SIGCSE Bulletin}} \bibinfo{volume}{34},
  \bibinfo{number}{2} (\bibinfo{year}{2002}).
\newblock


\bibitem[\protect\citeauthoryear{Cohoon, Wu, and Luo}{Cohoon
  et~al\mbox{.}}{2008}]%
        {cohoon08}
\bibfield{author}{\bibinfo{person}{J.M. Cohoon}, \bibinfo{person}{Z. Wu}, {and}
  \bibinfo{person}{L. Luo}.} \bibinfo{year}{2008}\natexlab{}.
\newblock \showarticletitle{Will {T}hey {S}tay or {W}ill {T}hey {G}o?}. In
  \bibinfo{booktitle}{\emph{Proceedings of the SIGCSE Conference}}.
\newblock
\showISBNx{978-1-59593-799-5}
\urldef\tempurl%
\url{https://doi.org/10.1145/1352135.1352273}
\showDOI{\tempurl}


\bibitem[\protect\citeauthoryear{Falkner, Szabo, Michell, Szorenyi, and
  Thyer}{Falkner et~al\mbox{.}}{2015}]%
        {falkner15}
\bibfield{author}{\bibinfo{person}{Katrina Falkner}, \bibinfo{person}{Claudia
  Szabo}, \bibinfo{person}{Dee Michell}, \bibinfo{person}{Anna Szorenyi}, {and}
  \bibinfo{person}{Shantel Thyer}.} \bibinfo{year}{2015}\natexlab{}.
\newblock \showarticletitle{Gender {G}ap in {A}cademia: {P}erceptions of
  {F}emale {C}omputer {S}cience {A}cademics}. In
  \bibinfo{booktitle}{\emph{Proceedings of the ITiCSE Conference}}.
\newblock
\showISBNx{978-1-4503-3440-2}
\urldef\tempurl%
\url{https://doi.org/10.1145/2729094.2742595}
\showDOI{\tempurl}


\bibitem[\protect\citeauthoryear{Frieze and Quesenberry}{Frieze and
  Quesenberry}{2013}]%
        {frieze13}
\bibfield{author}{\bibinfo{person}{Carol Frieze} {and}
  \bibinfo{person}{Jeria~L. Quesenberry}.} \bibinfo{year}{2013}\natexlab{}.
\newblock \showarticletitle{From {D}ifference to {D}iversity: {I}ncluding
  {W}omen in the {C}hanging {F}ace of {C}omputing}. In
  \bibinfo{booktitle}{\emph{Proceedings of the SIGCSE Conference}}.
\newblock
\showISBNx{978-1-4503-1868-6}
\urldef\tempurl%
\url{https://doi.org/10.1145/2445196.2445327}
\showDOI{\tempurl}


\bibitem[\protect\citeauthoryear{Goode}{Goode}{2008}]%
        {goode08}
\bibfield{author}{\bibinfo{person}{Joanna Goode}.}
  \bibinfo{year}{2008}\natexlab{}.
\newblock \showarticletitle{Increasing {D}iversity in {K}-12 {C}omputer
  {S}cience: {S}trategies from the {F}ield}. In
  \bibinfo{booktitle}{\emph{Proceedings of the SIGCSE Conference}}.
\newblock
\showISBNx{978-1-59593-799-5}
\urldef\tempurl%
\url{https://doi.org/10.1145/1352135.1352259}
\showDOI{\tempurl}


\bibitem[\protect\citeauthoryear{Goodman and Kruskal}{Goodman and
  Kruskal}{1954}]%
        {goodman54}
\bibfield{author}{\bibinfo{person}{Leo~A Goodman} {and}
  \bibinfo{person}{William~H Kruskal}.} \bibinfo{year}{1954}\natexlab{}.
\newblock \showarticletitle{Measures of association for cross classifications}.
\newblock \bibinfo{journal}{\emph{Journal of the American statistical
  association}} \bibinfo{volume}{49}, \bibinfo{number}{268}
  (\bibinfo{year}{1954}), \bibinfo{pages}{732--764}.
\newblock


\bibitem[\protect\citeauthoryear{Guzdial, Ericson, Mcklin, and
  Engelman}{Guzdial et~al\mbox{.}}{2014}]%
        {guzdial14}
\bibfield{author}{\bibinfo{person}{Mark Guzdial}, \bibinfo{person}{Barbara
  Ericson}, \bibinfo{person}{Tom Mcklin}, {and} \bibinfo{person}{Shelly
  Engelman}.} \bibinfo{year}{2014}\natexlab{}.
\newblock \showarticletitle{Georgia {C}omputes! {A}n {I}ntervention in a {U}{S}
  {S}tate, with {F}ormal and {I}nformal {E}ducation in a {P}olicy {C}ontext}.
\newblock \bibinfo{journal}{\emph{ACM Transaction on Computing Education
  (TOCE)}} \bibinfo{volume}{14}, \bibinfo{number}{2}, Article
  \bibinfo{articleno}{13} (\bibinfo{year}{2014}), \bibinfo{numpages}{29}~pages.
\newblock
\showISSN{1946-6226}
\urldef\tempurl%
\url{https://doi.org/10.1145/2602488}
\showDOI{\tempurl}


\bibitem[\protect\citeauthoryear{Habib, Ateeq, and Rehman}{Habib
  et~al\mbox{.}}{2014}]%
        {habib14}
\bibfield{author}{\bibinfo{person}{Hina Habib}, \bibinfo{person}{Muhammad
  Ateeq}, {and} \bibinfo{person}{Muzammil~UI Rehman}.}
  \bibinfo{year}{2014}\natexlab{}.
\newblock \showarticletitle{Motivational and {I}nfluential {F}actors for
  {C}hoice of {C}{S} {M}ajor: {A} {G}ender {A}ware {S}tudy}. In
  \bibinfo{booktitle}{\emph{Proceedings of the ICTLCE Conference}}.
\newblock


\bibitem[\protect\citeauthoryear{{Higher Education Research Institute \&
  Cooperative Institutional Research Program}}{{Higher Education Research
  Institute \& Cooperative Institutional Research Program}}{[n. d.]}]%
        {heri}
\bibfield{author}{\bibinfo{person}{{Higher Education Research Institute \&
  Cooperative Institutional Research Program}}.} \bibinfo{year}{[n.
  d.]}\natexlab{}.
\newblock \bibinfo{title}{{CIRP Freshman Survey}}.
\newblock
  \bibinfo{howpublished}{\nolinkurl{http://www.heri.ucla.edu/cirpoverview.php}}.
    (\bibinfo{year}{[n. d.]}).
\newblock


\bibitem[\protect\citeauthoryear{Ibe, Howsmon, Penney, Granor, DeLyser, and
  Wang}{Ibe et~al\mbox{.}}{2018}]%
        {ibe18}
\bibfield{author}{\bibinfo{person}{Nwannediya~Ada Ibe},
  \bibinfo{person}{Rebecca Howsmon}, \bibinfo{person}{Lauren Penney},
  \bibinfo{person}{Nathaniel Granor}, \bibinfo{person}{Leigh~Ann DeLyser},
  {and} \bibinfo{person}{Kevin Wang}.} \bibinfo{year}{2018}\natexlab{}.
\newblock \showarticletitle{Reflections of a Diversity, Equity, and Inclusion
  Working Group based on Data from a National CS Education Program}. In
  \bibinfo{booktitle}{\emph{Proceedings of the 49th ACM Technical Symposium on
  Computer Science Education}}. ACM, \bibinfo{pages}{711--716}.
\newblock


\bibitem[\protect\citeauthoryear{Klawe}{Klawe}{2013}]%
        {klawe13}
\bibfield{author}{\bibinfo{person}{Maria Klawe}.}
  \bibinfo{year}{2013}\natexlab{}.
\newblock \showarticletitle{Increasing {F}emale {P}articipation in {C}omputing:
  The {H}arvey {M}udd {C}ollege {S}tory}.
\newblock \bibinfo{journal}{\emph{Computer}} \bibinfo{volume}{46},
  \bibinfo{number}{3} (\bibinfo{year}{2013}).
\newblock


\bibitem[\protect\citeauthoryear{Knobelsdorf and Schulte}{Knobelsdorf and
  Schulte}{2007}]%
        {knobelsdorf07}
\bibfield{author}{\bibinfo{person}{Maria Knobelsdorf} {and}
  \bibinfo{person}{Carsten Schulte}.} \bibinfo{year}{2007}\natexlab{}.
\newblock \showarticletitle{Computer {S}cience in {C}ontext: {P}athways to
  {C}omputer {S}cience}. In \bibinfo{booktitle}{\emph{Proceedings of the Baltic
  Sea Conference on Computing Education Research}}.
\newblock
\showISBNx{978-1-920682-69-9}
\urldef\tempurl%
\url{http://dl.acm.org/citation.cfm?id=2449323.2449331}
\showURL{%
\tempurl}


\bibitem[\protect\citeauthoryear{Lishinski, Yadav, Good, and Enbody}{Lishinski
  et~al\mbox{.}}{2016}]%
        {lishinski16}
\bibfield{author}{\bibinfo{person}{Alex Lishinski}, \bibinfo{person}{Aman
  Yadav}, \bibinfo{person}{Jon Good}, {and} \bibinfo{person}{Richard Enbody}.}
  \bibinfo{year}{2016}\natexlab{}.
\newblock \showarticletitle{Learning to program: Gender differences and
  interactive effects of students' motivation, goals, and self-efficacy on
  performance}. In \bibinfo{booktitle}{\emph{Proceedings of the 2016 ACM
  Conference on International Computing Education Research}}. ACM,
  \bibinfo{pages}{211--220}.
\newblock


\bibitem[\protect\citeauthoryear{Margolis and Fisher}{Margolis and
  Fisher}{2003}]%
        {margolis03}
\bibfield{author}{\bibinfo{person}{Jane Margolis} {and} \bibinfo{person}{Allan
  Fisher}.} \bibinfo{year}{2003}\natexlab{}.
\newblock \bibinfo{booktitle}{\emph{Unlocking the {C}lubhouse: {W}omen in
  {C}omputing}}.
\newblock \bibinfo{publisher}{The MIT Press}.
\newblock


\bibitem[\protect\citeauthoryear{{National Center for Women and Information
  Technology}}{{National Center for Women and Information Technology}}{2014}]%
        {ncwit14}
\bibfield{author}{\bibinfo{person}{{National Center for Women and Information
  Technology}}.} \bibinfo{year}{2014}\natexlab{}.
\newblock \bibinfo{title}{{What is the Impact of Gender Diversity on Technology
  Business Performance? Research Summary}}.
\newblock \bibinfo{howpublished}{\nolinkurl{www.ncwit.org/businesscase}}.
  (\bibinfo{year}{2014}).
\newblock
\newblock
\shownote{Accessed: 2019-08-2.}


\bibitem[\protect\citeauthoryear{{National Center for Women and Information
  Technology}}{{National Center for Women and Information Technology}}{2019}]%
        {ncwit19}
\bibfield{author}{\bibinfo{person}{{National Center for Women and Information
  Technology}}.} \bibinfo{year}{2019}\natexlab{}.
\newblock \bibinfo{title}{{By the Numbers}}.
\newblock \bibinfo{howpublished}{\nolinkurl{www.ncwit.org/bythenumbers}}.
  (\bibinfo{year}{2019}).
\newblock
\newblock
\shownote{Accessed: 2019-08-2.}


\bibitem[\protect\citeauthoryear{{National Science Foundation}}{{National
  Science Foundation}}{[n. d.]a}]%
        {nsf12}
\bibfield{author}{\bibinfo{person}{{National Science Foundation}}.}
  \bibinfo{year}{[n. d.]}\natexlab{a}.
\newblock \bibinfo{title}{{Women, Minorities, and Persons with Disabilities in
  Science and Engineering}}.
\newblock
  \bibinfo{howpublished}{\nolinkurl{http://www.nsf.gov/statistics/2015/nsf15311/digest/}}.
    (\bibinfo{year}{[n. d.]}).
\newblock


\bibitem[\protect\citeauthoryear{{National Science Foundation}}{{National
  Science Foundation}}{[n. d.]b}]%
        {nsf17}
\bibfield{author}{\bibinfo{person}{{National Science Foundation}}.}
  \bibinfo{year}{[n. d.]}\natexlab{b}.
\newblock \bibinfo{title}{{Women, Minorities, and Persons with Disabilities in
  Science and Engineering}}.
\newblock
  \bibinfo{howpublished}{\nolinkurl{https://www.nsf.gov/statistics/2017/nsf17310/digest/fod-women/computer-sciences.cfm}}.
    (\bibinfo{year}{[n. d.]}).
\newblock


\bibitem[\protect\citeauthoryear{Pivkina, Pontelli, Jensen, and Haebe}{Pivkina
  et~al\mbox{.}}{2009}]%
        {pivkina09}
\bibfield{author}{\bibinfo{person}{Inna Pivkina}, \bibinfo{person}{Enrico
  Pontelli}, \bibinfo{person}{Rachel Jensen}, {and} \bibinfo{person}{Jessica
  Haebe}.} \bibinfo{year}{2009}\natexlab{}.
\newblock \showarticletitle{Young {W}omen in {C}omputing: {L}essons {L}earned
  from an {E}ducational \& {O}utreach {P}rogram}. In
  \bibinfo{booktitle}{\emph{Proceedings of the SIGCSE Conference}}.
\newblock
\showISBNx{978-1-60558-183-5}
\urldef\tempurl%
\url{https://doi.org/10.1145/1508865.1509042}
\showDOI{\tempurl}


\bibitem[\protect\citeauthoryear{Rheingans, D'Eramo, Diaz-Espinoza, and
  Ireland}{Rheingans et~al\mbox{.}}{2018}]%
        {rheingans18}
\bibfield{author}{\bibinfo{person}{Penny Rheingans}, \bibinfo{person}{Erica
  D'Eramo}, \bibinfo{person}{Crystal Diaz-Espinoza}, {and}
  \bibinfo{person}{Danyelle Ireland}.} \bibinfo{year}{2018}\natexlab{}.
\newblock \showarticletitle{A Model for Increasing Gender Diversity in
  Technology}. In \bibinfo{booktitle}{\emph{Proceedings of the 49th ACM
  Technical Symposium on Computer Science Education}}. ACM,
  \bibinfo{pages}{459--464}.
\newblock


\bibitem[\protect\citeauthoryear{Rorrer, Allen, and Zuo}{Rorrer
  et~al\mbox{.}}{2018}]%
        {rorrer18}
\bibfield{author}{\bibinfo{person}{Audrey~Smith Rorrer},
  \bibinfo{person}{Joseph Allen}, {and} \bibinfo{person}{Huifang Zuo}.}
  \bibinfo{year}{2018}\natexlab{}.
\newblock \showarticletitle{A national study of undergraduate research
  experiences in computing: Implications for culturally relevant pedagogy}. In
  \bibinfo{booktitle}{\emph{Proceedings of the 49th ACM Technical Symposium on
  Computer Science Education}}. ACM, \bibinfo{pages}{604--609}.
\newblock


\bibitem[\protect\citeauthoryear{Sahami and Piech}{Sahami and Piech}{2016}]%
        {sahami16}
\bibfield{author}{\bibinfo{person}{Mehran Sahami} {and} \bibinfo{person}{Chris
  Piech}.} \bibinfo{year}{2016}\natexlab{}.
\newblock \showarticletitle{As {C}{S} {E}nrollments {G}row, {A}re {W}e
  {A}ttracting {W}eaker {S}tudents?}. In \bibinfo{booktitle}{\emph{Proceedings
  of the ACM Technical Symposium on Computing Science Education}}.
\newblock


\bibitem[\protect\citeauthoryear{Sax}{Sax}{2014}]%
        {sax-ncge-2014}
\bibfield{author}{\bibinfo{person}{L. Sax}.} \bibinfo{year}{2014}\natexlab{}.
\newblock \bibinfo{title}{The {G}ender {G}ap in {S}{T}{E}{M}: {P}rogress and
  {C}hallenges between 1971-2011}.
\newblock \bibinfo{howpublished}{National Conference on Girls' Education
  (NCGE),
  \nolinkurl{http://www.ncgs.org/Pdfs/NCGE/2014/SessionC/TheGenderGapInSTEM.pdf}}.
    (\bibinfo{year}{2014}).
\newblock


\bibitem[\protect\citeauthoryear{Sax, Lehman, Jacobs, Kanny, Lim, Paulson, and
  Zimmerman}{Sax et~al\mbox{.}}{2015}]%
        {sax15}
\bibfield{author}{\bibinfo{person}{L. Sax}, \bibinfo{person}{K. Lehman},
  \bibinfo{person}{J. Jacobs}, \bibinfo{person}{A. Kanny}, \bibinfo{person}{G.
  Lim}, \bibinfo{person}{L. Paulson}, {and} \bibinfo{person}{H. Zimmerman}.}
  \bibinfo{year}{2015}\natexlab{}.
\newblock \showarticletitle{Anatomy of an {E}nduring {G}ender {G}ap: {T}he
  {E}volution of {W}omen's {P}articpation in {C}omputer {S}cience.}. In
  \bibinfo{booktitle}{\emph{Proceedings of the American Edcational Research
  Association}}.
\newblock


\bibitem[\protect\citeauthoryear{Sax}{Sax}{2008}]%
        {sax08}
\bibfield{author}{\bibinfo{person}{Linda~J Sax}.}
  \bibinfo{year}{2008}\natexlab{}.
\newblock \bibinfo{booktitle}{\emph{The {G}ender {G}ap in {C}ollege:
  {M}aximizing the {D}evelopmental {P}otential of {W}omen and {M}en.}}
\newblock \bibinfo{publisher}{Jossey-Bass}.
\newblock


\bibitem[\protect\citeauthoryear{Sax}{Sax}{2012}]%
        {sax12}
\bibfield{author}{\bibinfo{person}{Linda~J Sax}.}
  \bibinfo{year}{2012}\natexlab{}.
\newblock \showarticletitle{Examining the {U}nderrepresentation of {W}omen in
  {S}{T}{E}{M} {F}ields: {E}arly {F}indings from the {F}ield of {C}omputer
  {S}cience}.
\newblock \bibinfo{journal}{\emph{CSW Update Newsletter}}
  (\bibinfo{year}{2012}).
\newblock


\bibitem[\protect\citeauthoryear{Scutt, Gilmartin, Sheppard, and
  Brunhaver}{Scutt et~al\mbox{.}}{2013}]%
        {scutt13}
\bibfield{author}{\bibinfo{person}{M Scutt}, \bibinfo{person}{S Gilmartin},
  \bibinfo{person}{Sheri Sheppard}, {and} \bibinfo{person}{S Brunhaver}.}
  \bibinfo{year}{2013}\natexlab{}.
\newblock \showarticletitle{Research-informed {P}ractices for {I}nclusive
  {S}cience, {T}echnology, {E}ngineering, and {M}ath ({S}{T}{E}{M})
  {C}lassrooms: {S}trategies for {E}ducators to {C}lose the {G}ender {G}ap}.
\newblock \bibinfo{journal}{\emph{American Society of Engineering Education}}
  (\bibinfo{year}{2013}).
\newblock


\bibitem[\protect\citeauthoryear{Shapiro and Sax}{Shapiro and Sax}{[n. d.]}]%
        {shapiro11}
\bibfield{author}{\bibinfo{person}{Casey~A Shapiro} {and}
  \bibinfo{person}{Linda~J Sax}.} \bibinfo{year}{[n. d.]}\natexlab{}.
\newblock \showarticletitle{Major {S}election and {P}ersistence for {W}omen in
  {S}{T}{E}{M}}.
\newblock \bibinfo{journal}{\emph{New Directions for Institutional Research}}
  \bibinfo{volume}{2011} (\bibinfo{year}{[n. d.]}).
\newblock


\bibitem[\protect\citeauthoryear{{University of California Berkeley EECS
  Department}}{{University of California Berkeley EECS Department}}{[n. d.]}]%
        {berkeley}
\bibfield{author}{\bibinfo{person}{{University of California Berkeley EECS
  Department}}.} \bibinfo{year}{[n. d.]}\natexlab{}.
\newblock \bibinfo{title}{{Case Study -- A Plan of Action Measured}}.
\newblock
  \bibinfo{howpublished}{\nolinkurl{http://catalystsforchange.berkeley.edu/let-data-speak/plan-action-measured}}.
    (\bibinfo{year}{[n. d.]}).
\newblock


\bibitem[\protect\citeauthoryear{Vilner and Zur}{Vilner and Zur}{2006}]%
        {vilner06}
\bibfield{author}{\bibinfo{person}{Tamar Vilner} {and} \bibinfo{person}{Ela
  Zur}.} \bibinfo{year}{2006}\natexlab{}.
\newblock \showarticletitle{Once {S}he {M}akes {I}t, {S}he is {T}here: {G}ender
  {D}ifferences in {C}omputer {S}cience {S}tudy}. In
  \bibinfo{booktitle}{\emph{Proceedings of the ITiCSE Conference}}.
\newblock
\showISBNx{1-59593-055-8}
\urldef\tempurl%
\url{https://doi.org/10.1145/1140124.1140185}
\showDOI{\tempurl}


\bibitem[\protect\citeauthoryear{Wang, Hejazi~Moghadam, and
  Tiffany-Morales}{Wang et~al\mbox{.}}{2017}]%
        {wang17}
\bibfield{author}{\bibinfo{person}{Jennifer Wang}, \bibinfo{person}{Sepehr
  Hejazi~Moghadam}, {and} \bibinfo{person}{Juliet Tiffany-Morales}.}
  \bibinfo{year}{2017}\natexlab{}.
\newblock \showarticletitle{Social perceptions in computer science and
  implications for diverse students}. In \bibinfo{booktitle}{\emph{Proceedings
  of the 2017 ACM Conference on International Computing Education Research}}.
  ACM, \bibinfo{pages}{47--55}.
\newblock


\bibitem[\protect\citeauthoryear{Wood and Locke}{Wood and Locke}{1987}]%
        {wood87}
\bibfield{author}{\bibinfo{person}{Robert~E. Wood} {and}
  \bibinfo{person}{Edwin~A. Locke}.} \bibinfo{year}{1987}\natexlab{}.
\newblock \showarticletitle{The Relation of Self-Efficacy and Grade Goals to
  Academic Performance}.
\newblock \bibinfo{journal}{\emph{Educational and Psychological Measurement}}
  \bibinfo{volume}{47}, \bibinfo{number}{4} (\bibinfo{year}{1987}),
  \bibinfo{pages}{1013--1024}.
\newblock
\urldef\tempurl%
\url{https://doi.org/10.1177/0013164487474017}
\showDOI{\tempurl}


\end{thebibliography}

\end{document}